\documentclass{ws-ijmpd}
\usepackage[super,compress]{cite}
\usepackage{graphicx,psfig,epsf,epsfig}
\usepackage{hyperref}
\begin{document}
\markboth{Debashree Sen, and T.K. Jha}{Deconfinement of non-strange hadronic matter with nucleons and $\Delta$ baryons to quark matter in neutron stars}

\catchline{}{}{}{}{}
%
%%%%%%%%%%%%%%%%%%%%%%%%%%%%%%%%%%%%%%%%%%%%%%%%%%%%%%%%%%%%%%%%%%%%

\title{Deconfinement of non-strange hadronic matter with nucleons and $\Delta$ baryons to quark matter in neutron stars}

\author{Debashree Sen\footnote{p2013414@goa.bits-pilani.ac.in}}

\author{T.K. Jha\footnote{tkjha@goa.bits-pilani.ac.in}\\
}

\address{Birla Institute of Technology and Science-Pilani, K K Birla Goa Campus, NH-17B, Zuarinagar, Goa-403726, India}

\maketitle

\begin{history}
\received{Day Month Year}
\revised{Day Month Year}

\end{history}

\begin{abstract}
We explore the possibility of formation of $\Delta$ baryons (1232 MeV) in neutron star matter in an effective chiral model within the relativistic mean-field framework. With variation in delta-meson couplings, consistent with the constraints imposed on them, the resulting equation of state is obtained and the neutron star properties are calculated for static and spherical configuration. Within the framework of our model the critical densities of formation of $\Delta$s and the properties of neutron stars are found to be very sensitive to the iso-vector coupling compared to the scalar or vector couplings. We revisit the $\Delta$ puzzle and look for the possibility of phase transition from non-strange hadronic matter (including nucleons and $\Delta$s) to deconfined quark matter, based on QCD theories. The resultant hybrid star configurations satisfy the observational constraints on mass from the most massive pulsars PSR J1614-2230 and PSR J0348+0432 in static condition obtained with the general hydrostatic equilibrium based on GTR. Our radius estimates are well within the limits imposed from observational analysis of QLMBXs. The obtained values of $R_{1.4}$ are in agreement with the recent bounds specified from the observation of gravitational wave (GW170817)from binary neutron star merger. The constraint on baryonic mass from study of binary system PSR J0737-3039 is also satisfied with our hybrid equation of state.

\end{abstract}

\ccode{PACS numbers: 97.60.Jd, 26.60.+c, 26.60.−c, 25.75.Nq}

%\tableofcontents

\section{Introduction}
\label{intro}

At densities relevant to the core of neutron stars $\approx (5-10) \rho_0$, (where, $\rho_0 \sim 0.16~fm^{-3}$ is the normal nuclear matter density) the appearance and concentration of exotic matter like hyperons, $\Delta$ isobars, quarks and forms of bosonic condensates etc. present an exciting possibility \cite{Glen,Glen2,Glen3,Glen4,Glenq,Glen5,Glen6}. Of them, the $\Delta$ particles (1232 MeV) are of special interest because of the similarity in their quark structure to that of nucleons and therefore they can be treated on equal footing along the nucleons. Moreover, owing to low excitation energy ($m_{\Delta}-m_N$ = 293 MeV) and strong coupling with N-$\pi$ system, $\Delta$s often have various critical contributions to nuclear dynamics \cite{Cattapan}. They are formed in neutron star matter (NSM) only as metastable resonance states and like hyperons, the appearance of the $\Delta$ isobars depends strongly on the strength of coupling with mesons, which are largely unknown for the $\Delta$s at present. In this work we study the possibility of formation of $\Delta$ resonances in NSM and their effects on the structural properties of neutron stars (NS), with an effective chiral model for $\Delta$ couplings that are consistent with available literature. However, it is well known that the formation of such exotic matter like hyperons or $\Delta$ particles etc. in NSM softens the equation of state (EoS) considerably, reducing the maximum mass of the NS. This leads to the well-known hyperon and delta puzzles in light of the recent observational estimates of high mass pulsars like PSR J1614-2230 ($M = (1.928 \pm 0.017) M_{\odot}$) \cite{Fonseca} and PSR J0348+0432 ($M = (2.01 \pm 0.04) M_{\odot}$) \cite{Ant}. Many works have suggested various ways to deal with these additional degrees of freedom and to solve these puzzles. Refs. \cite{Kolomeitsev,Maslov,Zhu,Zhou,prc92,Torsten,del1,del4, DragoPRC,Chen,Lavagno,Hu,Oliveira,Oliveira2,Oliveira3,Rodrigues} have dealt with the delta puzzle and refs. \cite{Miyatsu,Stone,Dhiman,Dexheimer,Bednarek,Bednarek2, Weissenborn12,Weissenborn14,Agrawal,Lopes, Oertel,Colucci,Dalen,Lim,Rabhi,TKJ2,TKJ3,Sen} with the hyperon puzzle, in a phenomenological approach (using both relativistic and non-relativistic treatments) while works like \cite{Piarulli,Log,Schulze,Baldo,Vidana,Vidana2,Katayama,Yamamoto}  have been done within the microscopical framework 
for the same purpose. According to literature, the main mechanisms to solve such puzzles are (i) considering the effect of repulsive hyperon-hyperon interaction via exchange of vector mesons \cite{Bednarek,Weissenborn12,Oertel,Maslov} or scalar meson exchange \cite{Dalen}, (ii) inclusion of repulsive hyperonic three-body forces \cite{Vidana,Yamamoto,Takatsuka,Lonardoni}, (iii) effect of phase transition from hadronic to deconfined quark matter \cite{Ozel,Weissenborn11,Klahn,Bonanno,Lastowiecki,Dragoq1, Dragoq2,Bombaci16,Zdunik,Masuda16,Wu,Wu1}, (iv) calculating NS properties with modified/extended theories of gravity \cite{Doneva,Brax,Capozziello,Arapoglu,Astashenok}. In this work we address the delta puzzle and adopt the third possible mechanism i.e., hadron-quark phase transition to solve it.

 Like many works \cite{prc92,Zhou,Oliveira3,Hu,Chen}, here we do not include the strange baryons in the pure hadronic part of NSM for simplicity. $\Delta$ particles have been earlier analyzed in various field theoretical approaches with varying coupling strengths \cite{del1,del4}. The appearance/disappearance of the $\Delta$ baryons in NSM, depends on their coupling strength with the mesons \cite{DragoPRC,Hu,DragoPRD}. Refs. \cite{prc92,DragoPRC} also show that critical densities of $\Delta$s in NSM may be as low as 2$\rho_0$ and that such early appearances are strongly co-related to the slope parameter of the symmetry energy. However, in absence of any conclusive experimental data for the potential depth of $\Delta$s in normal nuclear matter, the $\Delta$-meson couplings are poorly known. Refs. \cite{Riek,DragoPRC} with the references therein suggest that from the studies of electron-nucleus scattering \cite{Wehrberger,Koch,Connell}, photoabsorption \cite{Alberico} and pion-nucleus scattering \cite{Horikawa,Nakamura}, the $\Delta$ potential $V_{\Delta}$ is shallow, attractive and within the range -30 MeV + $V_N$ $\leq V_{\Delta} \leq V_N$ in normal nuclear matter. However, refs. \cite{Kolomeitsev,Maslov} suggest the range of $\Delta$ potential to be $V_{\Delta}=-(50-100)$ MeV. Owing to similar quark structure as the nucleons, refs. \cite{Mosozkowski,Garpman} have considered the $\Delta$-meson coupling strengths to be same as that of nucleons (universal coupling scheme). But ref. \cite{Bohr} has shown reduction in Gamow-Teller transition strengths due to coupling with $\Delta$s, suggesting weaker coupling strengths for $\Delta$s compared to that of nucleons. On the other hand, the finite density QCD sum-rule calculations \cite{qcd1} predict larger scalar $\Delta$ coupling than those for the nucleons while the corresponding vector coupling for $\Delta$s to be two times smaller than those of the nucleons. However, refs. \cite{qcd,Lavagno,del1} show that the $\Delta$ couplings must be based on the following criteria : (i) the second minimum of the energy per baryon must lie above the saturation energy of normal nuclear matter, (ii) there are no $\Delta$ isobars present at the saturation density and (iii) the scalar field is more or same attractive while the vector potential is less or same repulsive for $\Delta$s compared to that of nucleons. A possible range for the choice of scalar and vector couplings for the $\Delta$s, satisfying the above requirements, is provided in form of a triangle in ref. \cite{qcd}. The couplings suggested by QCD sum-rule predictions \cite{qcd1} do not fall within the estimates derived on the basis of the above three rules. Refs. \cite{Wehrberger,Zhu,qcd} also suggest that in Hartree approximation, the difference between the scalar and vector $\Delta$-meson couplings is $\approx$ 0.2. However, ref. \cite{Zhang} very recently concluded that the cross sections considered are very less sensitive to both the scalar and vector couplings. But it is to be noted that the appearance and concentration of $\Delta$ particles in NSM and the resulting NS properties are very sensitive to the $\Delta$-$\rho$ meson coupling strength and symmetry energy \cite{prc92,Torsten}. Thus for overall lack of any conclusive experimental data, we vary the $\Delta$ couplings moderately and investigate their effects on the resulting EoS and NS properties. We also test the sensitivity of the appearance and concentration of $\Delta$s and the resultant NS properties on the scalar, vector and iso-vector couplings individually. For this purpose we span the allowed range of the scalar and vector couplings prescribed in ref. \cite{qcd}.

 In the present work the effects of presence of $\Delta$ baryons in NSM are studied with an effective chiral model \cite{TKJ,TKJ2,TKJ3,TKJ4} and the results are compared and correlated with the findings of other models, used for the same purpose. The model has already been tested for NS properties using hyperon rich matter only \cite{TKJ2,TKJ3}. It embodies chiral symmetry, where the mass of the nucleons and the mesons are dynamically generated \cite{TKJ}. It has also been emphasized that the non-linear chiral model may mimic the effective three body forces, which can have decisive roles to play at high densities \cite{three}. The chosen model parameter set for this work, is on the basis of reasonable nuclear matter saturation properties. It is also well constrained and related to the vacuum expectation value of the scalar field. Hence there are very few free parameters available to adjust the saturation properties \cite{TKJ}.

 As mentioned earlier, we choose hadron-quark phase transition as a mechanism to solve the delta puzzle. A huge amount of work is done over the years to understand the  possible presence of quark matter and hadron-quark mixed phases in the NS core. The hadron-quark phase boundary is not so prominently marked but is supposed to be co-existing \cite{Glen,Glenq}. Of the various models considered in literature to explain the properties of pure quark matter and mixed phases, in the present work, we consider a simple form of MIT bag model \cite{Chodos} with corrections due to strong repulsive interactions in the thermodynamic quark potential \cite{Alford05,Schramm,Fraga,Weissenborn11,Bombaci17} to describe unpaired quark phase. First order correction due to strong interaction \cite{Fraga} and perturbative effects \cite{Kapusta,Benhar,Nakazato,Uechi,Bombaci16,Bombaci17} may also be considered but several works \cite{Steiner,Prakash,Yazdizadeh,Burgio,Miyatsu,Liu} emphasize that the effects of perturbative corrections can also be realized by changing the bag constant. The mixed phase properties like critical density for appearance of quarks, the density range over which mixed phase can extend and the EoS etc. are governed by the charge neutrality condition between the two phases viz. the global charge neutrality condition i.e., the Gibbs construction (GC) \cite{Glen,Glenq,Orsaria,Rotondo} or the local charge neutrality condition i.e., the Maxwell construction (MC) \cite{Bhattacharya,LogBom}. As pointed out in refs. \cite{Maruyama1,Maruyama2} that GC corresponds to an unrealistic scenario of zero surface tension and Coulomb energy between hadron and quark phases. As a result it becomes quite unfavorable in terms of energy and relevant density range of neutron stars. Moreover, GC may yield large NS mass due to the “masquerade” effect \cite{Alford05}. Under such circumstances, MC provides a more physically justified and relevant way of describing the properties of hybrid NS. In this work we compare the results of both the constructions to study the properties of hybrid NS. The value of bag constant and the interaction strength also play a very important role in determining the phase transition properties as well as the structural properties of NS \cite{Bag,Li,Yudin,Logoteta2}.

 The present manuscript is planned as follows. After discussion on the model attributes in section 2, we present the formalism employed to include the $\Delta$ resonances in NSM, phase transition to quark matter and the mixed phase properties in section II, finally culminating in the resulting NS properties in section 3. The sensitivity of appearance and concentration of $\Delta$s and the resultant NS properties to individual delta-meson couplings are shown in \ref{app_sensitivity}. We finally conclude in the closing section.
 
\section{Formalism}
\label{sec:1}
\subsection{The Effective Chiral Model with nucleons and $\Delta$ baryons}
 The Lagrangian of effective chiral model is given by eq. \ref{Lagrangian}. The detailed attributes of model for nuclear matter can be found in \cite{TKJ,TKJ2,TKJ3,TKJ4}. However, for the sake of completeness we describe the basic ingredients of the model.
%%%%%%%%%%%%%%%%%%%%%%%%%%%%%
\begin{eqnarray}
\mathcal{L} &=& \overline{\psi}_B \Biggl[ \left(i \gamma_{\mu} \partial^{\mu} - g_{\omega B}~ \gamma_{\mu} \omega^{\mu} 
-\frac{1}{2} g_{\rho B}~ \overrightarrow{\rho_{\mu}} \cdot \overrightarrow{\tau} \gamma^{\mu} \right)-g_{\sigma B} 
\left(\sigma + i \gamma_5 \overrightarrow{\tau} \cdot \overrightarrow{\pi} \right) \Biggr] \psi_B \nonumber \\
&+& \frac{1}{2} \left(\partial_{\mu} \overrightarrow{\pi} \cdot \partial^{\mu} \overrightarrow{\pi} 
+ \partial_{\mu} \sigma ~ \partial^{\mu} \sigma \right) -{\frac{\lambda}{4}} \left(x^2-x_0^2\right)^2 
- \frac{\lambda B}{6} (x^2-x_0^2)^3 - \frac{\lambda C}{8}(x^2-x_0^2)^4 \nonumber \\
&-& \frac{1}{4}F_{\mu\nu}F^{\mu\nu} 
+\frac{1}{2}\sum_B {g_{\omega B}}^2~x^2~\omega_\mu \omega^\mu 
- \frac{1}{4}~\overrightarrow{R_{\mu\nu}} \cdot 
\overrightarrow{R^{\mu\nu}}+\frac{1}{2}~m_\rho^2 ~\overrightarrow{\rho_\mu} \cdot \overrightarrow{\rho^\mu} 
\protect\label{Lagrangian}
\end{eqnarray}

In the model, the baryon spinor $\psi_B$ interacts via exchange of scalar meson $\sigma$, the vector meson $\omega$ (783 MeV), the iso-vector $\rho$-meson (770 MeV) with respective couplings $g_{\sigma B}$, $g_{\omega B}$ and $g_{\rho B}$. The sumover index $B$ in eq. \ref{Lagrangian} signifies all the baryonic states including the $\Delta$s (sumover index $B=n,p,\Delta^{-,0,+,++}$). Here, $x^2 = ({\pi}^2+\sigma^2$) makes the $\sigma$ and $\omega$ fields chiral invariant and the higher order scalar field invariant terms are taken accordingly with $B$ and $C$ as the coupling constants. The spontaneous breaking of the chiral symmetry at ground state lends mass to the baryons ($m_B$), the scalar meson ($m_{\sigma}$) and the vector meson ($m_{\omega}$), in terms of the expectation value of the scalar condensate $<\sigma_0> = <x_0>$, and are given by

\begin{eqnarray}
m_B = g_{{\sigma_B}} x_0,~~ m_{\sigma} = \sqrt{2\lambda} x_0,~~
m_{\omega} = g_{{\omega_N}} x_0~.
\end{eqnarray}
\noindent

 The mean field treatment makes $<\pi>=0$ and the pion mass $m_\pi = 0$. Hence we neglect their explicit contributions. The equation of motion of the vector field ($\omega$), the scalar field ($\sigma$) (in terms of $Y=x/x_0 = {m_B}^{\star}/{m_B}$) and the iso-vector field ($\rho$) are respectively given by

\begin{eqnarray}
\omega_0 = \dfrac {\sum\limits_{B} g_{\omega_B} \rho_B}{\Bigl(\sum\limits_{B} g_{\omega_B}^2 \Bigr) x^2}
\protect\label{vector_eq}
\end{eqnarray}

\begin{eqnarray} 
\hspace*{-1.0cm}\sum_B \Biggl[(1-Y^2)-\frac{B}{C_{\omega_N}}(1-Y^2)^2+\frac{C}{C_{\omega_N}^2}(1-Y^2)^3 +2\frac{C_{\sigma_B}~C_{\omega_N}}{m_B^2 ~Y^4} \dfrac{\Bigl(\sum\limits_{B} g_{\omega_B} \rho_B\Bigr)^2}{\sum\limits_{B} {g_{\omega_B}}^2} - 2 \sum_B \frac{~C_{\sigma_B}~\rho_{SB}}{m_B~ Y} \Biggr]=0 \nonumber \\
\protect\label{scalar_eq}
\end{eqnarray} 

\begin{eqnarray}
\rho_{03} = \sum_{B} \frac{{g_{\rho}}_B}{m_\rho^2} I_{3B}~\rho_{B}.
\protect\label{isovector_eq}
\end{eqnarray}

where, $\omega_0$ and $\rho_{03}$ are the mean field approximate or vacuum expectation values of the $\omega$ and the $\rho$ fields, respectively. $m_{\rho}$ and $I_{3B}$ are the mass of $\rho$ meson and isospin third component of each baryon species, respectively. The quantities $\rho_B$ and $\rho_{SB}$ are the vector and the scalar densities, respectively of each baryon species. Therefore in terms of Fermi momenta $k_B$, 

\begin{eqnarray} 
\rho_{SB}=\frac{\gamma_B}{2 \pi^2} \int^{k_B}_0 dk ~k^2 \frac{m_B^*}{\sqrt{k^2 + {m^{*}_{B}}^2}}
\end{eqnarray}

and the total baryon density ($\rho$) is

\begin{eqnarray} 
\rho = \sum_B \rho_B =\frac{1}{2\pi^2} \sum_{B} \gamma_B \int^{k_B}_0 dk ~k^2  
\end{eqnarray}

$\gamma_B$ is the spin degeneracy factor of the nucleons and $\Delta$s. The total energy density $\varepsilon$ and pressure $P$ are 

\begin{eqnarray} 
\hspace*{-0.5cm}\varepsilon &=& \frac{m_B^2}{8~C_{\sigma_B}}(1-Y^2)^2-\frac{m_B^2 B}{12~C_{\omega_N}C_{\sigma_B}}(1-Y^2)^3
+\frac{C m_B^2}{16 ~C_{\omega_N}^2~ C_{\sigma_B}}(1-Y^2)^4 +\frac{1}{2Y^2}C_{\omega_N} \dfrac {\Bigl(\sum\limits_{B} g_{\omega_B} \rho_B\Bigr)^2}{\sum\limits_{B} {g_{\omega_B}}^2} \nonumber \\
&+& \frac{1}{2}~m_\rho^2 ~\rho_{03}^2 + \frac{1}{\pi^2} \sum_B \gamma_B \int_{0}^{k_B} k^2 \sqrt{(k^2+{m_B^*}^2)} ~dk 
+ \frac{\gamma}{2\pi^2} \sum_{\lambda= e,\mu^-} \int_{0}^{k_\lambda} k^2 \sqrt{(k^2+{m_\lambda}^2)}~ dk
\protect\label{EoS1}
\end{eqnarray}

\begin{eqnarray}         
\hspace*{-1.2cm}P &=& -\frac{m_B^2}{8~C_{\sigma_B}}(1-Y^2)^2+\frac{m_B^2 B}{12~C_{\omega_N}~C_{\sigma_B}}(1-Y^2)^3
-\frac{C~ m_B^2}{16~ C_{\omega_N}^2 C_{\sigma_B}}(1-Y^2)^4
+\frac{1}{2Y^2}~C_{\omega_N} \dfrac {\Bigl(\sum\limits_{B} g_{\omega_B} \rho_B\Bigr)^2}{\sum\limits_{B} {g_{\omega_B}}^2} \nonumber \\ 
&+& \frac{1}{2}~m_\rho^2 ~\rho_{03}^2 + \frac{1}{3\pi^2} \sum_B \gamma_B \int_{0}^{k_B} \frac{k^4}{ \sqrt{(k^2+{m_B^*}^2)}}~ dk 
+ \frac{\gamma}{6\pi^2} \sum_{\lambda= e,\mu^-} \int_{0}^{k_\lambda} \frac{k^4}{ \sqrt{(k^2+{m_\lambda}^2)}}~ dk
\protect\label{EoS2} 
\end{eqnarray}

 In the equations above, $Y=m^*_B/m_B$ and $C_{iB}=(g_{iB}/m_i)^2$, where $i = \sigma, \omega, \rho$ while $C_{\omega_N}={1}/{x_0^2}$.
 
 The isospin triplet $\rho$ mesons are incorporated to account for the asymmetric nuclear matter. Although it is possible to consider the effect of interaction of the $\rho$ mesons with the scalar and the pseudoscalar mesons similar to the $\omega$ meson and to dynamically generate the mass of $\rho$ mesons similar to that of the scalar and vector mesons, we choose to consider an explicit mass term for the iso-vector $\rho$ meson similar to what was considered in \cite{Sahu1,Sahu2,TKJ2,TKJ3,TKJ,TKJ4}. The coupling strength for the $\rho$ meson is obtained by fixing the symmetry energy coefficient $J = 32$ MeV at $\rho_0$ and is given by,

\begin{equation}
J = \frac{C_{\rho_N} k_N^3}{12\pi^2} + \frac{k_N^2}{6\sqrt{(k_N^2 + m^{\star 2})}}
\end{equation}

where $C_{\rho_N} \equiv g^2_{\rho_N}/m^2_{\rho}$ and $k_B=(6\pi^2 \rho_B/{\gamma})^{1/3}$.

 We construct NSM with the baryons ($n, p, \Delta^{-,0,+,++}$) and the leptons ($e, \mu$). As mentioned earlier, the similarity in the substructure and comparable mass between nucleon and $\Delta$s enables one to treat the later on equal footing with the nucleons. The relevant strong processes are then, 

\begin{eqnarray}
n + n &\longrightarrow& \Delta^{-} + p \\\nonumber
p + n &\longrightarrow& \Delta^{0} + p \\\nonumber
n + p &\longrightarrow& \Delta^{+} + n \\\nonumber
p + p &\longrightarrow& \Delta^{++} + n.
\end{eqnarray}

However, for a stable configuration one needs to impose the required conditions of charge neutrality and chemical equilibrium conditions, which are as follows :

\begin{eqnarray} 
\sum_{B} Q_B~\rho_B + \sum_{l} Q_l~\rho_l = 0 
\protect\label{charge_neutrality}
\end{eqnarray}

\begin{eqnarray} 
\mu_B=\mu_n-Q_B \mu_e \nonumber \\
\mu_\mu=\mu_e
\protect\label{chemeq}
\end{eqnarray}

In the equations above, $Q_B$, $\rho_B$ and $Q_l$, $\rho_l$ are the charge states and density of the baryons and leptons, respectively and $\mu_n$ and $\mu_e$ are chemical potentials of neutron and electron, respectively. 

The corresponding baryon chemical potential is given by
\begin{eqnarray} 
\mu_B=\sqrt{{k_B}^2+{m^*_B}^2} ~+~g_{{\omega}_B} ~\omega_0 ~ +~g_{{\rho}_B} I_{3B}\rho_{03}
\protect\label{bar_chem}
\end{eqnarray} 

 At high enough momentum (density), when the nucleon chemical potential reaches the mass state of the $\Delta$ baryons, the latter start appearing in dense matter, guided by the charge neutrality condition. Similarly, $\mu$ appears at the expense of the electrons. For a given momentum/density and coupling strengths between the $\Delta$s and the mesons, the energy density and pressure of many body system is evaluated using eqs. \ref{EoS1} and \ref{EoS2}. 
 
 One needs to specify the coupling strengths for the $\Delta$ resonances ($x_{\sigma_\Delta}=g_{\sigma_\Delta}/g_{\sigma_N},~ x_{\omega_\Delta}=g_{\omega_\Delta}/g_{\omega_N},~ x_{\rho_\Delta}=g_{\rho_\Delta}/g_{\rho_N}$), which are largely unknown. For our current work, we look into the variation of the same (to be discussed in the result section) according to that prescribed in \cite{qcd} and study their effects on the NS composition and structure. The potential depth for the $\Delta$s ($U_{\Delta}$) at saturation density can be calculated as

\begin{eqnarray} 
U_{\Delta}=x_{\sigma_\Delta}m_N (Y-1) + x_{\omega_\Delta}C_{\omega_N}\rho_0
\protect\label{potdSNM}
\end{eqnarray} 

and when $\rho$ mesons are incorporated, the $\Delta$ potential is given as

\begin{eqnarray} 
U_{\Delta}=x_{\sigma_\Delta}m_N (Y-1) + x_{\omega_\Delta}C_{\omega_N}\rho_0 - \frac{1}{2}x_{\rho}I_{3\Delta}C_{\rho_N}\rho_0
\protect\label{potdPNM}
\end{eqnarray} 

where, $I_{3\Delta}$ is the third component of isospin of the $\Delta$ particles.

\subsection{The model parameter}
%\label{sec:2.1}
The parameter set of the effective chiral model for the present work is chosen from ref. \cite{TKJ} and is listed in table \ref{table-1}, along with the saturation properties. It is  obtained self-consistently by fixing the standard state properties at T $= 0$ for symmetric nuclear matter in the mean-field analysis \cite{TKJ2},\cite{param1},\cite{param2}.

\begin{table}[ht!]
\tbl{Parameters of the nuclear matter models considered for the present work (adopted from \cite{TKJ}) are displayed. Listed are the saturation properties such as binding energy per nucleon $B/A$, nucleon effective mass $m^{\star}_N$/$m_N$, the symmetry energy coefficient $J$, slope parameter ($L_0$) and the nuclear matter incompressibility ($K$) all defined at saturation density $\rho_0$. $C_{\sigma_N}$, $C_{\omega_N}$ and $C_{\rho_N}$ are the corresponding scalar, vector and iso-vector couplings. $B$ and $C$ are the higher order couplings of the scalar field. The scalar meson mass $m_{\sigma}$ is also displayed.}
%\hline
{\begin{tabular}{@{}cccccccccccccc@{}} \toprule
Model & $C_{\sigma_N}$ & $C_{\omega_N}$ & $C_{\rho_N}$ & $B/m^2$ & $C/m^4$ & $m_N^{\star}/m_N$ & $m_{\sigma}$ & $f_{\pi}$ & $K$ & $B/A$ & $J(L_0)$ & $\rho_0$ \\ &  $fm^2$ & $fm^2$ & $fm^2$ & $fm^2$ & $fm^4$ & & $MeV$ & $MeV$ & $MeV$ & $MeV$ & $MeV$ & $fm^{-3}$ \\ \colrule
NM-I\hphantom{0}  & 6.772\hphantom{0}  & 1.995\hphantom{0}  & 5.285\hphantom{0} & -4.274\hphantom{0}   & 0.292\hphantom{0}    & 0.85\hphantom{0}  & 509.644\hphantom{0}   & 139.710\hphantom{0}  & 303\hphantom{0}  & -16.3\hphantom{0}   & 32(89)\hphantom{0}  & 0.153\hphantom{0} \\ \botrule
\label{table-1}
\end{tabular}}
\end{table}

 It is to be noted that the nucleon effective mass ($m^{\star}_N=0.85~ m_N$) in the present case is quite high in comparison to other relativistic mean field models \cite{rmf,rmf1}, which has got large impact on the resulting EoS of dense matter and NS properties. Further, one of the salient features of the model is that at high density, the nucleon effective mass increases \cite{TKJ} which is unlike any other rmf models considered in the literature. This effect leads to softer EoS at high density. Therefore it would be interesting to investigate the effects of phase transition at high densities. The nuclear incompressibility ($K = 303$~ MeV) is in good agreement with findings from \cite{k1,k2,Stone2}. The corresponding symmetry energy coefficient for the model is $J = 32$~MeV which is in excellent agreement with the limits imposed from the empirical and experimental findings \cite{j1}. However, the slope parameter for the present model comes out to be $L_0 = 87$~MeV, which is a bit larger than the recent limits imposed \cite{l1}. Ref.\cite{rmf1} suggests the value of $L_0 = (25 - 115)$~MeV though. The rest of the saturation properties have the standard values such as the saturation density ($\rho_0 = 0.153$~$\rm{fm^{-3}}$) and binding energy per particle ($B/A = -16.3$~MeV) for SNM. The EoS for both SNM and PNM with this parameter set are also in good agreement with the heavy-ion collision data \cite{hic} as shown in \cite{TKJ}.

\subsection{Hadron-Quark Phase transition and Phase Equilibrium}
Unpaired quark matter (u, d and s) with electrons is an exotic system. To understand the properties of such a system, we take into account a simple form of MIT bag model, characterized by the bag constant $B$ and strong interaction strength $\alpha_4$ between the quarks \cite{Alford05,Fraga,Schramm,Bombaci17}. The MIT Bag model is based on QCD theories \cite{Chodos,Weissenborn11} and the thermodynamic quark potential \cite{Weissenborn11} is given by

\begin{eqnarray} 
\Omega_{QM} =  \sum_{i=u,d,s,e} \Omega_i + \frac{3\mu^4}{4\pi^2} (1 - \alpha_4) + B
\protect\label{quarkpot}
\end{eqnarray} 

where, $\Omega_i$ is the grand potential of different quark species and electrons and $\mu$ is the baryon chemical potential.

 We consider first order phase transition from hadronic phase to quark phase and the EoS of the resultant hybrid star can be obtained using the well-known Gibbs Construction (GC) or Maxwell Construction (MC).

\subsubsection{Gibbs Construction (GC)}

 The Gibbs criteria \cite{Glen,Glenq,Li,Logoteta2,Orsaria,Rotondo,Bhattacharya} for the global charge neutrality condition states that the total mixed phase must be charge neutral and it is given by

\begin{eqnarray}
\chi\rho_c^Q + (1-\chi)\rho_c^H + \rho_c^l =0
\protect\label{charge_neutrality_mpGC}
\end{eqnarray}
where, $\rho_c^Q$,~$\rho_c^H$ and $\rho_c^l$ are the total charge densities of quarks, hadrons and leptons, respectively and the volume fraction for quark is given by $0\leq\chi\leq1$. Then $\chi=0~ \& ~1$ denotes pure hadronic and pure quark phases, respectively. The pressure and chemical potentials of the two phases are related by the following equations

\begin{eqnarray}
P_{MP}=P_H(\mu_B,\mu_e)=P_Q(\mu_B,\mu_e)
\protect\label{eos_P_mp}
\end{eqnarray}
and
\begin{eqnarray}
\mu_B^H = \mu_B^Q
\protect\label{mub_mp}
\end{eqnarray}

\vspace*{-1cm}

\begin{eqnarray}
\mu_e^H = \mu_e^Q
\protect\label{mue_mp}
\end{eqnarray}

The energy density and baryon density of the mixed phase are respectively given by

\begin{eqnarray}
\varepsilon_{MP}=\chi\varepsilon_Q + (1-\chi)\varepsilon_H
\protect\label{eos_e_mp}
\end{eqnarray}

and

\begin{eqnarray}
\rho_{MP}=\chi\rho_Q + (1-\chi)\rho_H
\protect\label{eos_rho_mp}
\end{eqnarray}

\subsubsection{Maxwell Construction (MC)}

 In case of Maxwell construction \cite{Glen,Bhattacharya,LogBom,Schramm} the hadron and quark phases are in direct contact with each other. $\mu_B$ is continuous while there is jump in $\mu_e$ at the interface between the two phases. The pressure remains constant in the density interval of phase transition in case of MC unlike that of GC. Therefore with MC, the pressure and chemical potentials are given by eqs. \ref{eos_P_mp} and \ref{mub_mp}. The charge neutrality conditions in MC, called the local charge neutrality condition, states that unlike GC the individual hadron and quark phase must be charge neutral. They are given as given as
 
\begin{eqnarray}
q_H(\mu_B,\mu_e) = 0 ~;~ q_Q(\mu_B,\mu_e) = 0
\protect\label{charge_neutrality_mpMC}
\end{eqnarray} 

 The crust part of the NS has a much low density and this part is taken care by using the BPS EoS \cite{BPS} in the EoS. 

\subsection{Neutron Star Structure \& Properties}
The equations for the structure of a relativistic spherical and static star composed of a perfect fluid were derived from Einstein's equations by Tolman \cite{tov}, Oppenheimer and Volkoff \cite{tov1}, which are

\begin{equation}
\frac{dP}{dr}=-\frac{G}{r}\frac{\left[\varepsilon+P\right ]
\left[M+4\pi r^3 P\right ]}{(r-2 GM)},
\label{tov1}
\end{equation}

\begin{equation}
\frac{dM}{dr}= 4\pi r^2 \varepsilon,
\label{tov2}
\end{equation}
\noindent

with $G$ as the gravitational constant and $M(r)$ as the enclosed gravitational mass. We have used $c=1$. 
For the specified EoS, these equations can be integrated from the origin for a given choice of central energy density $(\varepsilon_c)$. The value of radius $r~(=R)$, where the pressure vanishes defines the surface of the star.

 The baryonic mass $M_B(r)$ of neutron star is defined as the total baryon number enclosed in the volume of radius $R$ multiplied by atomic mass unit (a.m.u.). The difference between the gravitational mass and the baryonic mass gives the gravitational binding of the star.
 
\begin{equation} 
M_B(r)=\int_{0}^{R} 4\pi r^2 ~\varepsilon~ m_B \left(1 - \frac{2GM}{r}\right)^{1/2} dr
\label{barmass}
\end{equation}

where, $m_B$ is the mass of baryon.

\section{Result and Discussions}
\subsection{Neutron star with $\Delta$ baryons}
$\Delta$s appear in dense matter at the expense of nucleons. However, one needs to fix their interaction strengths ($x_{\sigma_\Delta}=g_{\sigma_\Delta}/g_{\sigma_N}=x_{\sigma},~ x_{\omega_\Delta}=g_{\omega_\Delta}/g_{\omega_N}=x_{\omega},~ x_{\rho_\Delta}=g_{\rho_\Delta}/g_{\rho_N}=x_{\rho}$) with the respective mesons. In accordance with the finite density QCD sum-rule calculations \cite{qcd,qcd1}, we fix the scalar \& vector coupling for $\Delta$s. The QCD sum-rule calculations indicate that in comparison to the nucleons, the vector coupling strength for $\Delta$s are considerably smaller than that of the scalar counterpart i.e., $x_{\sigma} \geq 1$ and $x_{\omega} \leq 1$ \cite{qcd1,Lavagno}. Therefore, we fix $x_{\sigma} = 1.35$ and $x_{\omega} = 1.0$ as the first prescription (CS-I and CS-II). However, in the absence of any conclusive experimental data or theoretical constraint on $x_{\rho}$, we take two cases where, $x_{\rho}$ = 1.0 (CS-I and CS-III)~\&~ 0.5 (CS-II and CS-IV), respectively. 
As the second prescription, we fix $x_{\sigma} = 1.20$ and $x_{\omega} = 0.8$ (CS-III and CS-IV) and $x_{\rho} =$ 1.0 (CS-I and CS-III) \& 0.5 (CS-III and CS-IV) as before.
We are also restricted for the choice of $x_{\rho}$ coupling since we find that the $\Delta$ baryons do not appear in NSM with further increase in the $\rho$ coupling ($x_{\rho} > 1$) for the present model considered. The present coupling scheme ($x_{\sigma},x_{\omega}$) : ($1.35,1.0$), ($1.2,0.8$)) thus respects the systematics of $\Delta$ self-energy as indicated by refs. \cite{qcd,qcd1}.

\begin{center}
CS-I ~:   $x_{\sigma}=1.35$, $x_{\omega}=1.0$, $x_{\rho} = 1.0$ \\
CS-II ~:  $x_{\sigma}=1.35$, $x_{\omega}=1.0$, $x_{\rho} = 0.5$ \\
CS-III : $x_{\sigma}=1.20$, $x_{\omega}=0.8$, $x_{\rho} = 1.0$ \\
CS-IV :  $x_{\sigma}=1.20$, $x_{\omega}=0.8$, $x_{\rho} = 0.5$
\end{center}

The above prescriptions for $\Delta$ scalar and vector couplings are within the acceptable range ref.\cite{qcd1}.

\begin{figure}[!ht]
\centering
\includegraphics[width=0.48\textwidth]{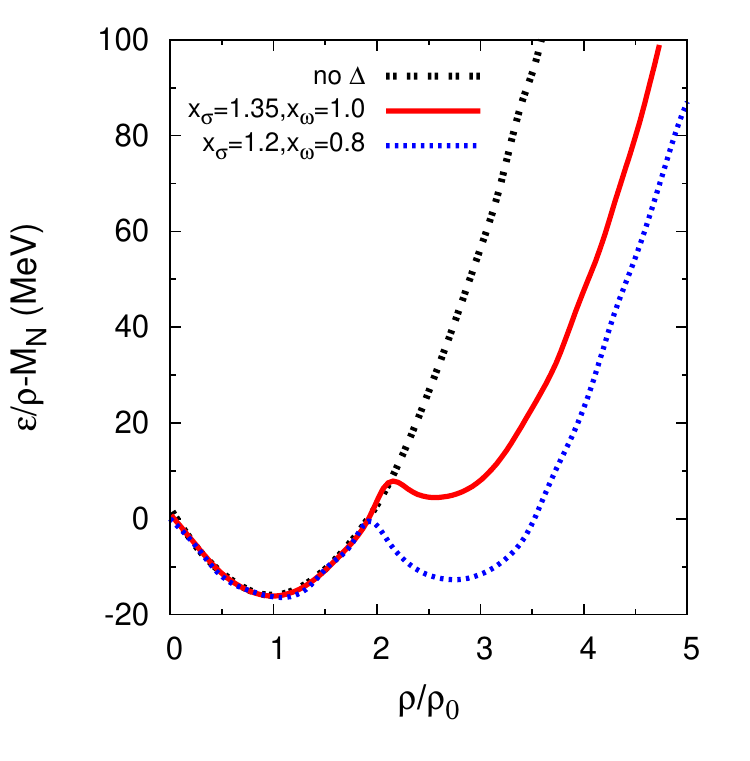}
\caption{Binding energy per baryon versus normalized baryon density for nuclear matter with and without including $\Delta$ isobars for the different coupling schemes.}
\protect\label{ebd}
\end{figure}

 In fig. \ref{ebd} we plot the energy per particle as a function of the normalized baryon density.

We find that the second minima of the energy per baryon lie well above the saturation energy of normal nuclear matter (-16.3~MeV) considered in this work. The second minima lie approximately at 8~MeV (2.5$\rho_0$) for $x_{\sigma} = 1.35, x_{\omega} = 1.0$  and at -14~MeV (2.8$\rho_0$) for $x_{\sigma} = 1.2, x_{\omega} = 0.8$. This indicates that $\Delta$s can exist with nucleons in NSM only as metastable resonance states \cite{qcd1,Lavagno}. There is not much difference in the binding energies for the two set of couplings considered because the decrease in binding energy with decrease of $x_{\omega}$ \cite{qcd,Oliveira3} is counterbalanced by the effect of increase in binding energy with the decrease of $x_{\sigma}$. Similar result is found in \cite{Oliveira3,Lavagno}. The corresponding $\Delta$ potentials for the two cases are are -110~MeV and -102.5~MeV, respectively which are quite large compared to the suggested shallow attractive potential depths in \cite{Riek},\cite{DragoPRC}. However, the large values of the $\Delta$ potentials obtained in the present work lie within the range suggested by the data available \cite{Kolomeitsev,Maslov}. Our choice of couplings $x_{\sigma}\geq 1$ and $x_{\omega}\leq 1$ also makes the scalar field more/same attractive while the vector potential less/same repulsive for $\Delta$s than for nucleons, thereby meeting the third criterion for $\Delta$ couplings \cite{qcd1}.

\begin{figure}[!ht]
\centering
\includegraphics[width=0.48\textwidth]{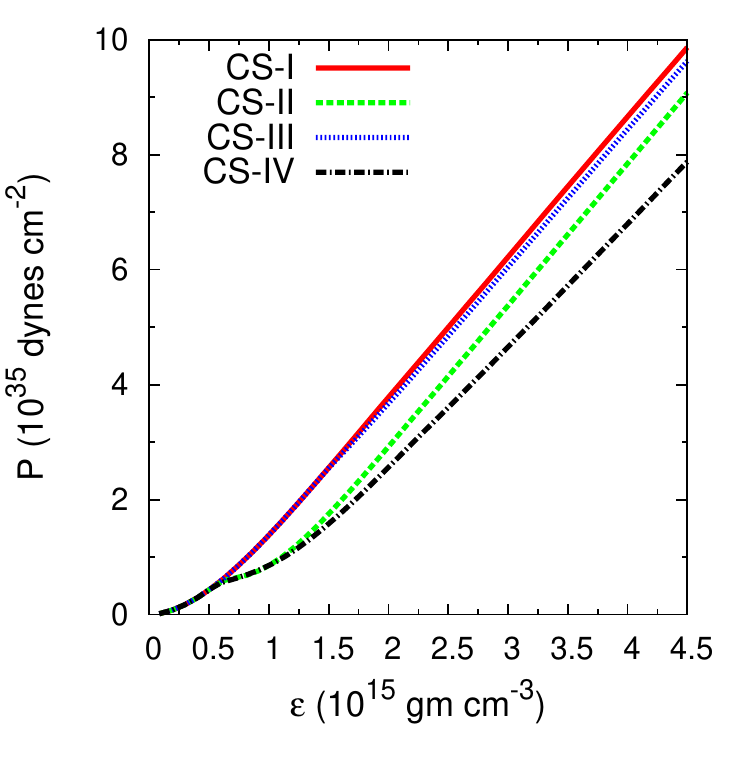}
\caption{Equation of State ($\varepsilon ~vs.~ P$) for neutron star matter including $\Delta$ resonances for the different coupling schemes as described in the text.}
\protect\label{eos_d}
\end{figure}

 The corresponding EoS ($\varepsilon ~vs.~ P$) is plotted in fig. \ref{eos_d} for the four coupling schemes including $\Delta$ baryons. 
 
\begin{figure}[!ht]
\centering
\includegraphics[width=0.5\textwidth]{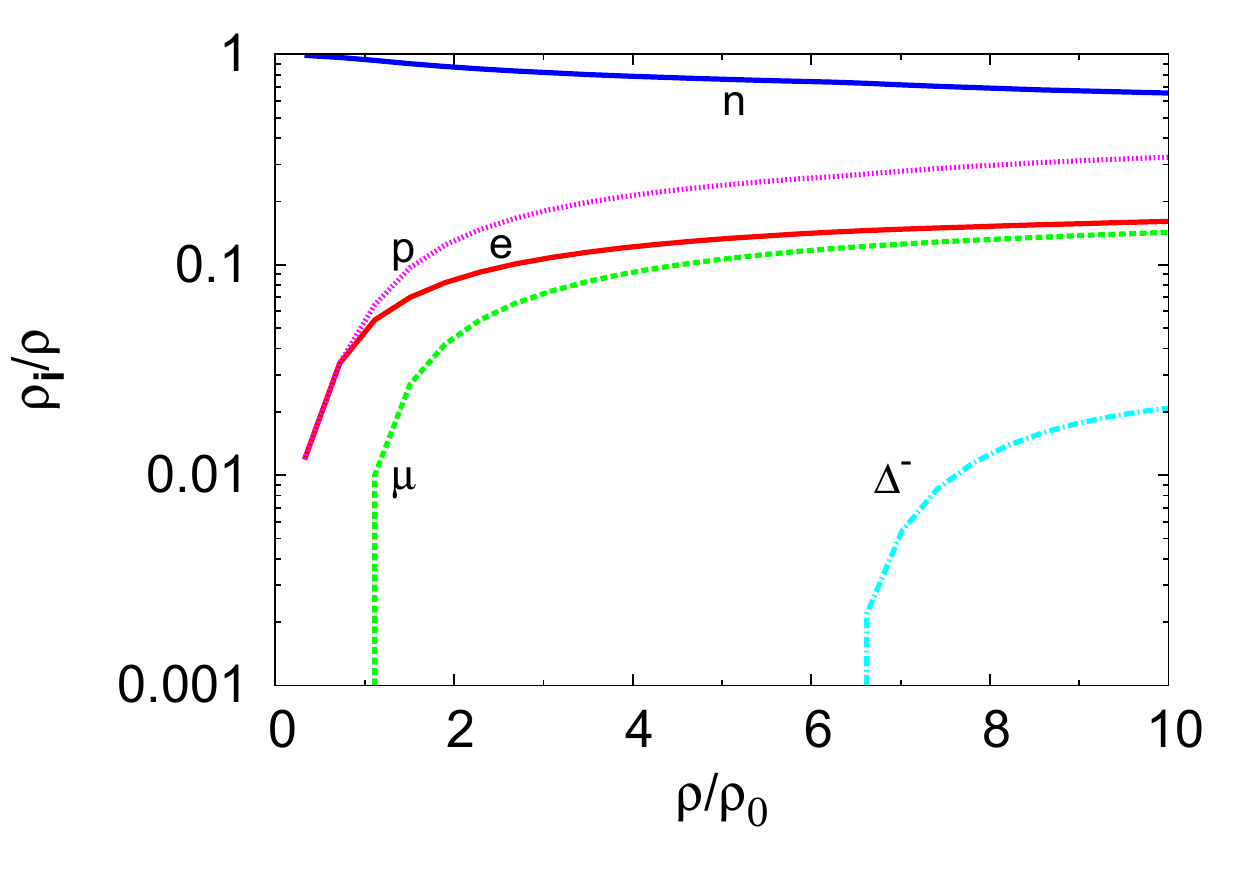}
\caption{\it Non-strange particles in neutron star matter with coupling schemes CS-I.}
\protect\label{pf-1}
\end{figure}

\begin{figure}[!ht]
\centering
\includegraphics[width=0.5\textwidth]{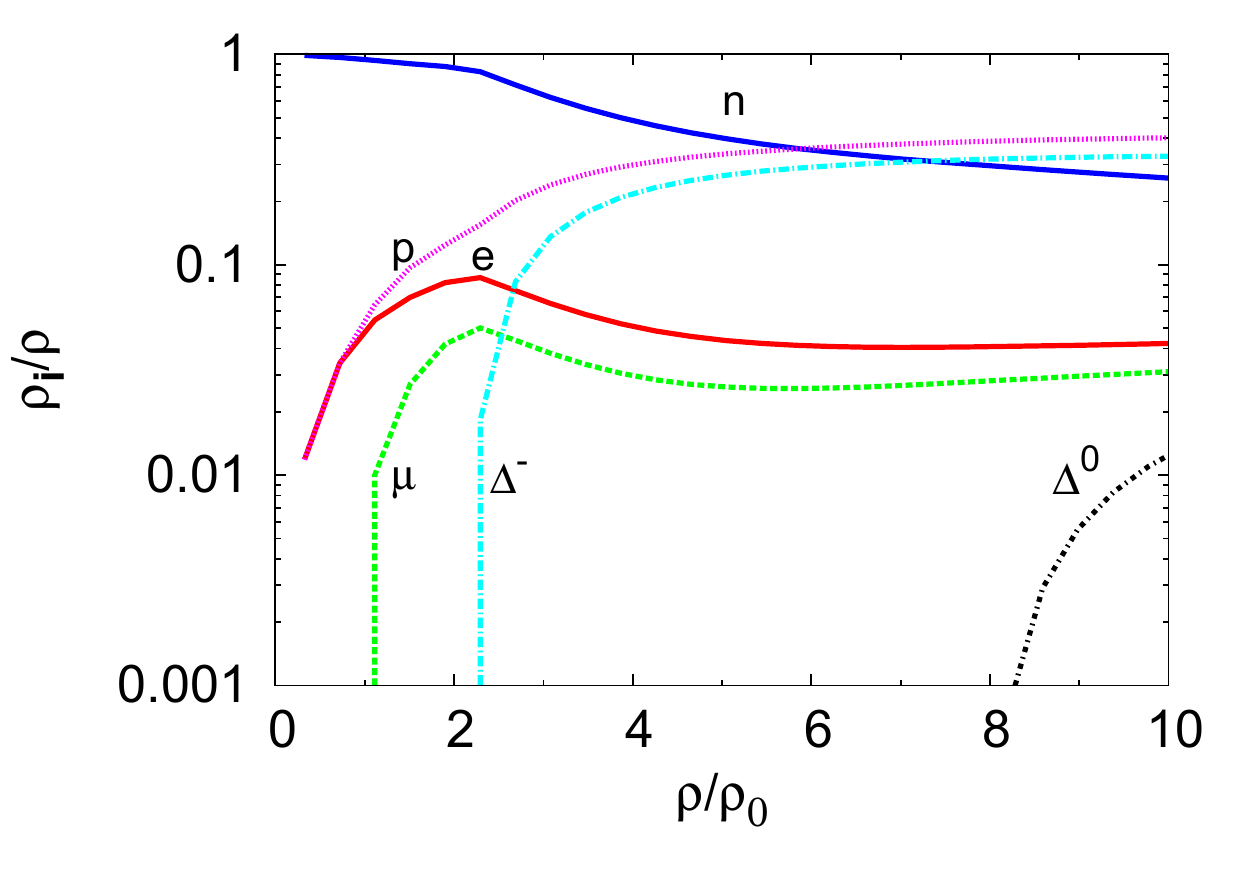}
\caption{\it Non-strange particles in neutron star matter with coupling schemes CS-II.}
\protect\label{pf-2}
\end{figure}

\begin{figure}[!ht]
\centering
\includegraphics[width=0.5\textwidth]{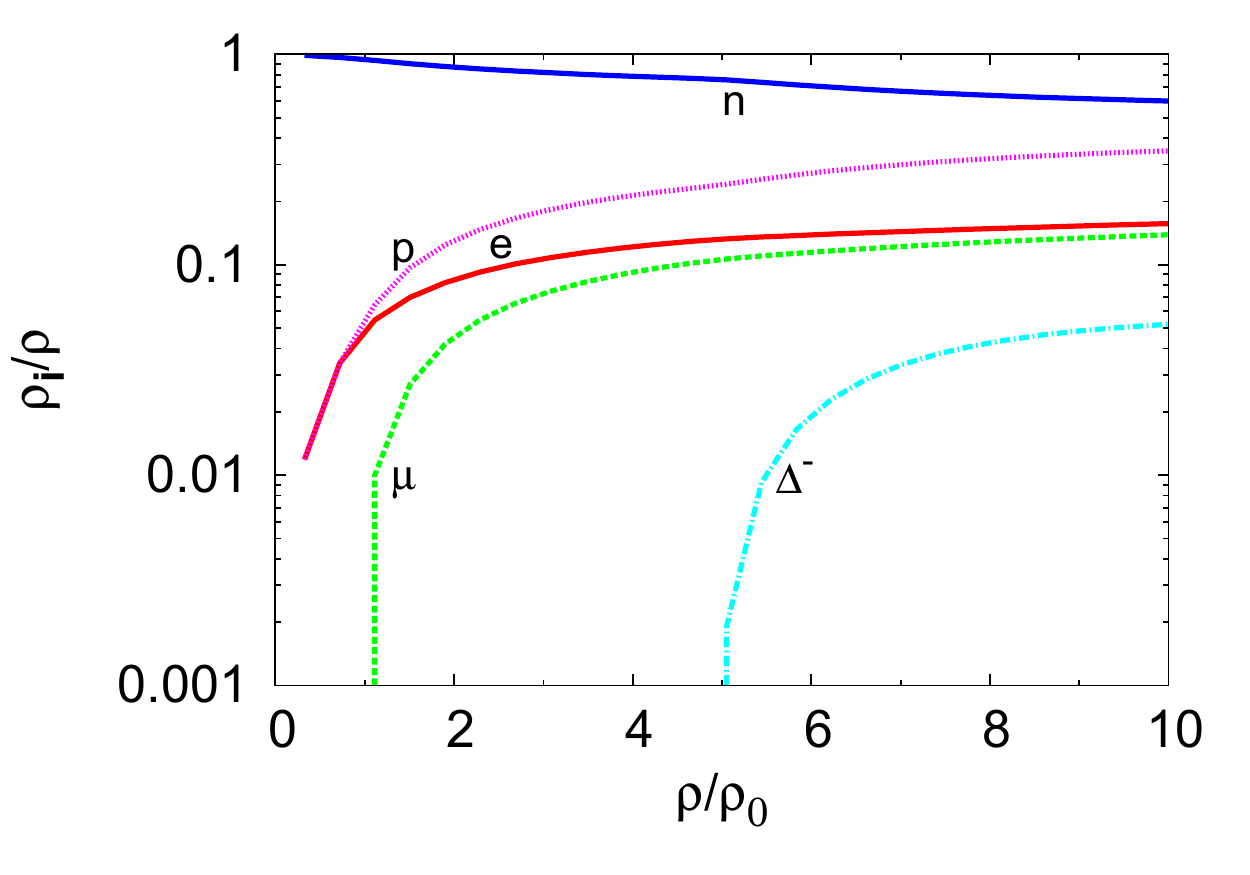}
\caption{\it Non-strange particles in neutron star matter with coupling schemes CS-III.}
\protect\label{pf-3}
\end{figure}

\begin{figure}[!ht]
\centering
\includegraphics[width=0.5\textwidth]{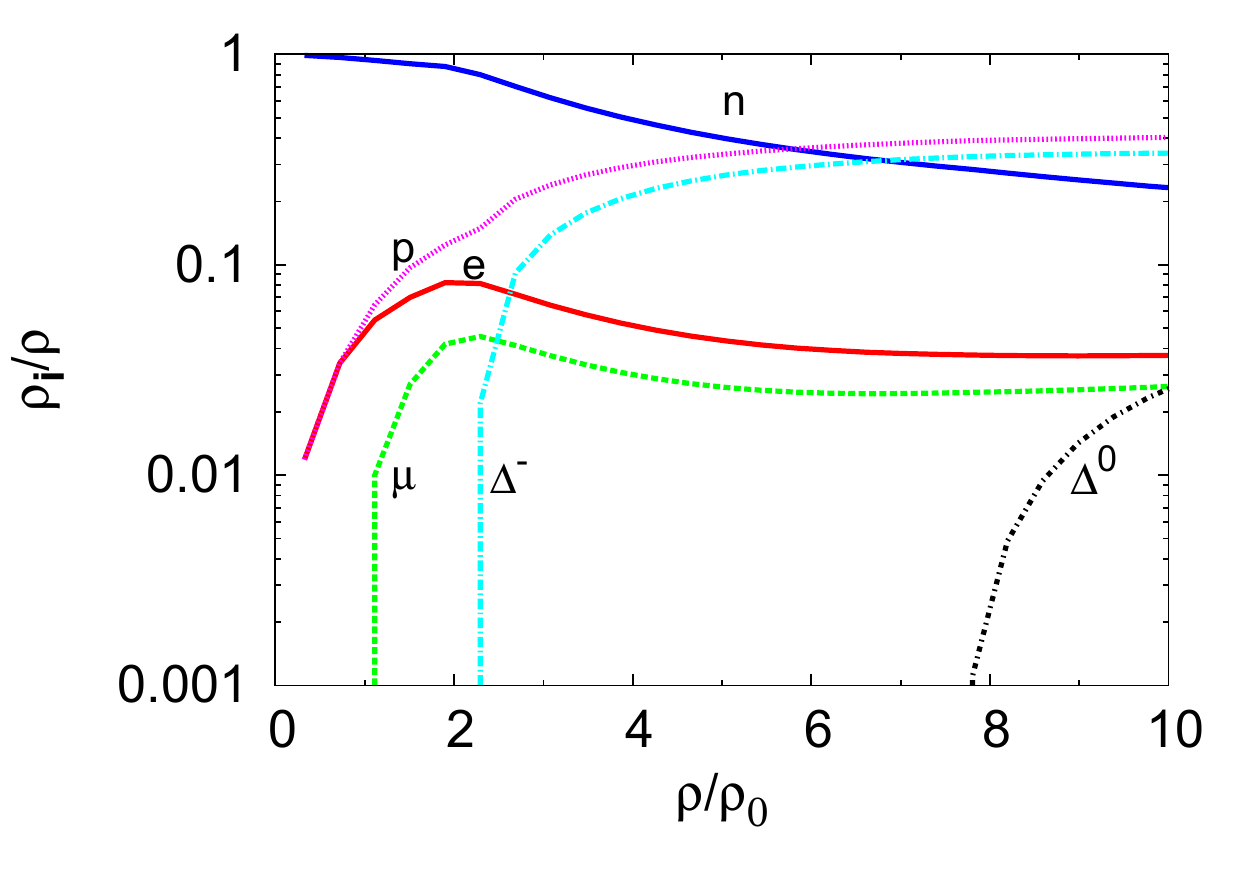}
\caption{\it Non-strange particles in neutron star matter with coupling schemes CS-IV.}
\protect\label{pf-4}
\end{figure}

 It is to be noted that the effect of strength of the iso-vector coupling is more pronounced when $x_{\sigma} = 1$. With decrease in the value of $x_{\rho}$, both neutron and electron chemical potentials drop. Therefore there is considerable softening of EoS with increased value of $x_{\rho}$. For NSM, the neutron and electron chemical potentials along with charge neutrality condition decide the relative appearance and population of different particles in the matter as shown in figs. \ref{pf-1}, \ref{pf-2}, \ref{pf-3} \& \ref{pf-4} for the different coupling schemes.

 None of the coupling schemes predicts the presence of $\Delta^+$ and $\Delta^{++}$ at relevant densities due to rapid exhaustion of neutron chemical potential. As expected, the negatively charged particles are the favored species over neutral and positively charged particles. In order to maintain charge neutrality of the matter the appearance of $\Delta^-$ is compensated by the enhanced presence of protons for smaller value of $x_{\rho}$ for all the cases. With decrease in the $\rho$ meson coupling strength the $\Delta$s start to appear at much lower density ($\approx2\rho_0$) and are also a major constituent in the matter along with the nucleons. The early appearance and their concentration  makes the EoS comparatively softer, resulting in lower maximum mass of NS so obtained. Similar feature has also been noted in \cite{DragoPRC}. We will discuss this result subsequently. For all the variations in couplings taken, $\Delta$s appear at $\approx$ 2$\rho_0$ when $x_{\rho} = 0.5$ (CS-II \& CS-IV). Their appearance shifts to higher densities with higher $x_{\rho} = 1$ (6.5$\rho_0$ for CS-I and 5$\rho_0$ for CS-III). $\Delta^0$ starts appearing at $\approx 8\rho_0$ only if the $\rho$ meson coupling is lowered (CS-I and CS-III). We find that deleptonization happens faster with smaller value of $x_{\rho}$ as leptons are used up to maintain the charge neutrality of matter. However, none other than the $\Delta^-$ appears in the dense matter till $10\rho_0$ for $x_{\rho} = 1$. The appearance of $\Delta$ particles in matter shifts to higher densities and their concentration in dense matter also decreases with increase in the $\rho$ meson coupling.

\begin{figure}[!ht]
\centering
\includegraphics[width=0.48\textwidth]{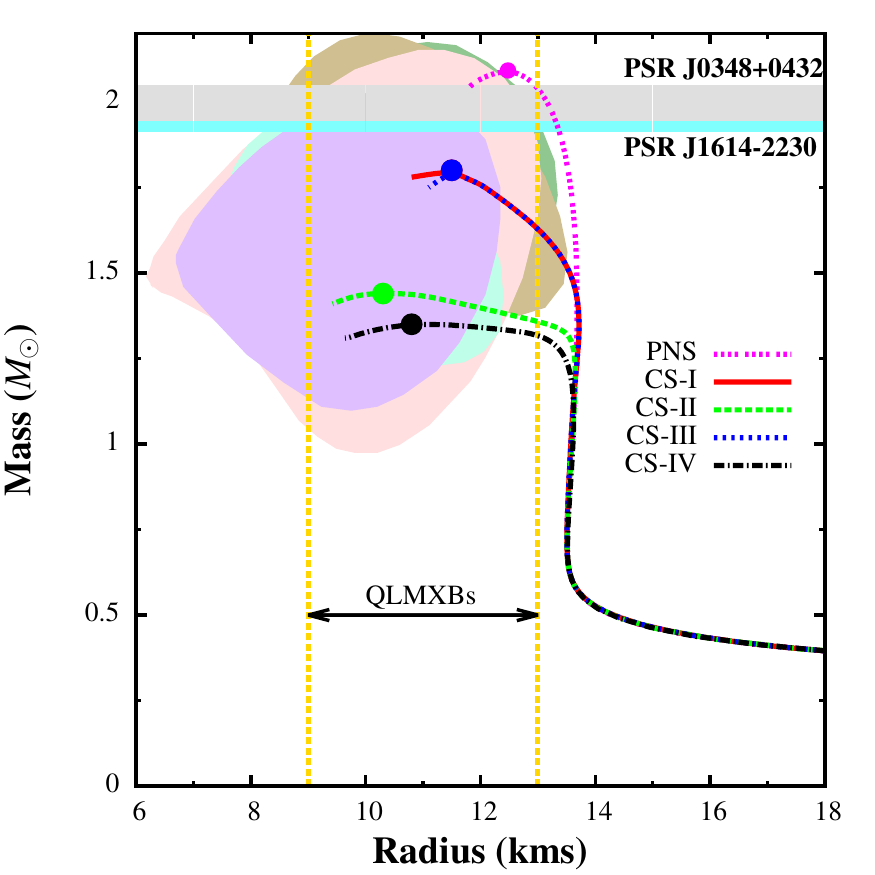}
\caption{Gravitational mass and radius relationship for pure nucleon star (PNS) and neutron stars with $\Delta$ baryons for different coupling schemes prescribed in the text. The maximum mass limits imposed from recent observation of high mass stars PSR J1614-2230 ($M=(1.928 \pm 0.017) M_{\odot}$) \cite{Fonseca} and PSR J0348+0432 ($M=(2.01 \pm 0.04) M_{\odot}$) \cite{Ant} are also indicated. The region (marked with arrow) includes the estimation of radius within $(9 - 13)~km$ from spectral analysis of quiescent X-ray transients in low mass binaries (QLMXBs) \cite{fortin}. The areas enclosed by the dark green, golden, pink, aquamarine and purple curves indicate radius constraints from \cite{Ozel2} for sources 4U1820-30, 4U1724-207, KS1731-260, EXO1745-248 and 4U1608-52, respectively.}
\protect\label{mrd}
\end{figure}

 In fig. \ref{mrd} we plot the mass-radius relationship with the obtained EoSs.
 
  For the static NS configurations (fig. \ref{mrd}), we find that due to formation of $\Delta$s, there is considerable reduction in maximum mass of NS compared to the maximum mass ($M = 2.1 ~M_{\odot}$) obtained with pure nucleonic matter. The maximum mass is largest for $x_{\rho} = 1$ (CS-I and CS-III) among the four coupling schemes considered. The mass drops down when the $\rho$ coupling is reduced. The masses obtained for $x_\rho = 1$ with the considered coupling schemes is $1.80~M_{\odot}$ and the corresponding radius is in the range 11.5~km. For $x_{\rho} = 0.5$ (CS-II and CS-IV), the mass and radius of the star are $(1.44~ \& ~1.35)~M_{\odot}$ and $(10.3 - 10.8)$~km, respectively. None of the current models satisfy maximum mass constraints of $M \approx 2 ~M_{\odot}$ \cite{Fonseca,Ant} from recent observations.  However, with the inclusion of $\Delta$s, the radius obtained is quite small, consistent with the results of ref. \cite{Torsten}. Moreover, the obtained values of $R_{1.4}$ are in agreement with the recent bounds specified from the observation of gravitational wave (GW170817) from binary neutron star (BNS) merger \cite{Abbot,Fattoyev,Most,De}. Our prediction of NS radius for all the couplings is in excellent agreement with the observational analysis of QLMBXs \cite{fortin} and in general for canonical NS configurations \cite{latt}. The maximum mass and corresponding radius predicted by the models considered satisfy the radius constraint of \cite{Ozel2}. The central density of the star is approximately $7.5\rho_0$ for all couplings taken for the static case. The results from hydrostatic equilibrium conditions viz. the central density of the star $\rho_c$, the maximum gravitational mass $M$ (in $M_{\odot}$), the baryonic mass $M_B$ (in $M_{\odot}$) and the radius $R$ (in km) in static and spherical configurations are listed in table \ref{table-2} for the different coupling schemes.
 
\begin{table*}[ht!]
\centering
\tbl{Critical densities for appearance of $\Delta^-$ ($\rho_{\Delta^-}^{crit}$) and $\Delta^0$ ($\rho_{\Delta^0}^{crit}$) and static neutron star properties for variation in the coupling schemes ie., $x_{i_\Delta} = g_{i_\Delta}/g_{i_N} = x_i$, where $i=\sigma, \omega, \rho$. The results from hydrostatic equilibrium conditions such as the central density of the star $\rho_c$ ($\rho_0$), the gravitational mass $M$ (in $M_{\odot}$), the baryonic mass $M_B$ (in $M_{\odot}$), radius $R$ (km) and $R_{1.4}$ (km) are tabulated.}
{\scriptsize{
\setlength{\tabcolsep}{4.5pt}
\begin{tabular}{cccccccccccccc}
\hline
\hline
\multicolumn{1}{c}{Coupling Scheme}&
\multicolumn{1}{c}{$x_{\sigma}$}&
\multicolumn{1}{c}{$x_{\omega}$} &
\multicolumn{1}{c}{$x_{\rho}$} &
\multicolumn{1}{c}{$U_{\Delta^-}$} &
\multicolumn{1}{c}{$\rho_{\Delta^-}^{crit}$} &
\multicolumn{1}{c}{$\rho_{\Delta^0}^{crit}$} &
\multicolumn{1}{c}{$\rho_c$} &
\multicolumn{1}{c}{$M$} &
\multicolumn{1}{c}{$M_{B}$} & 
\multicolumn{1}{c}{$R$} & 
\multicolumn{1}{c}{$R_{1.4}$} & \\
\multicolumn{1}{c}{} &
\multicolumn{1}{c}{} &
\multicolumn{1}{c}{} &
\multicolumn{1}{c}{} &
\multicolumn{1}{c}{(MeV)} &
\multicolumn{1}{c}{($\rho_0$)} &
\multicolumn{1}{c}{($\rho_0$)} &
\multicolumn{1}{c}{($\rho_0$)} &
\multicolumn{1}{c}{($M_{\odot}$)} &
\multicolumn{1}{c}{($M_{\odot}$)} &
\multicolumn{1}{c}{($km$)} & 
\multicolumn{1}{c}{($km$)} & \\
\hline
CS-I   &1.35   &1.0   &1.0  &-10.5 &6.5   &-    &7.8 &1.80  &1.89  &11.5 &13.4\\
CS-II  &       &      &0.5  &-70.5 &2.1   &8.1  &7.0 &1.44  &1.47  &10.3 &12.1\\ 
CS-III &1.20   &0.8   &1.0  &-1.2   &5.0   &-    &7.6 &1.80  &1.88  &11.5 &13.4\\
CS-IV  &      &       &0.5  &-61.2 &2.1   &7.9  &7.4 &1.35  &1.39  &10.8 &-\\ 
\hline
\hline
\end{tabular}
}}
%\end{center}
\protect\label{table-2}
\end{table*} 
 
 Overall, the appearance and concentration of the $\Delta$s are found to be very sensitive to the $x_{\rho}$ coupling as can be seen from figs. \ref{pf-1},\ref{pf-2},\ref{pf-3} and \ref{pf-4}. We find that with increasing $x_{\rho}$, there is a considerable shift to higher densities where $\Delta^-$ appears. This is also consistent with the findings of ref. \cite{Zhu}. For $x_{\rho} > 1$, we do not find any exotic species in matter at relevant densities. It is interesting to find that with different choice of scalar and vector coupling, the critical densities at which they appear do not change appreciably for smaller value of $x_{\rho}$. It can be seen that for lower $x_{\rho}$, the EoS becomes softer (fig. \ref{eos_d}) and leads to lower mass and radius for the NS (fig. \ref{mrd}). The difference in the global properties of the star as well as the composition of matter is negligible when we keep the scalar \& vector coupling in the same ratio. Therefore we test the sensitivity of formation of $\Delta$s in NSM \& resulting NS properties to the variation of individual scalar, vector \& iso-vector couplings in \ref{app_sensitivity}.

 As seen from fig. \ref{mrd}, the maximum mass constraint ($M \approx 2 ~M_{\odot}$) \cite{Ant} is not satisfied only with nucleons and $\Delta$. Therefore we now look for possible phase transition from hadronic matter to unpaired quark matter.

\subsection{Hadron-Quark Phase transition in Neutron Star}
We consider the u and d quark masses to be negligible ($\approx$ 0~MeV) compared to the mass of the s quark ($m_s$=100~MeV) \cite{Nakazato,PDG}. Refs. \cite{Steiner,Prakash,Yazdizadeh,Burgio,Miyatsu,Liu} suggest that the perturbative effects can also be realized by changing the bag constant $B$ and interaction strength $\alpha_4$. These values really play very important roles in determining the phase transition properties as well as the structural properties of NS and it is well known that increase in these values gives stiffer EoS, yielding more massive NS \cite{Bag,Li,Yudin,Logoteta2}. However, the limits to the values of $B$ are still not properly known. The value of $B$ used in the literature ranges from $\sim ((100 - 300)$ MeV)$^4$ ~\cite{Steiner,Buballa,Novikov,Baym}~. Moreover, \cite{Benhar,Satz} state that lattice calculations suggest the range to be $\sim 210$~MeV/fm$^3$. According to GW170817 observation and measurement of tidal deformability $\Lambda_{1.4}$ \& radius $R_{1.4}$, a recent work \cite{EnPingZhou} suggests that $B^{1/4} = (134.1 - 141.4)$ MeV and $\alpha_4 = (0.56 - 0.91)$ for a low-spin prior while for the high-spin priors $B^{1/4} = (126.1 - 141.4)$ MeV and $\alpha_4 = (0.45 - 0.91)$ considering pure quark stars while \cite{Nandi} suggests the maximum values of $B$ and $\alpha_4$ for hybrid stars with three RMF models like NL3, TM1 and NL3$\omega\rho$ to describe the hadronic part of the star. Therefore to understand the properties of hybrid stars, we therefore choose two moderate values of both $B$ and $\alpha_4$ as ($B^{1/4}$ MeV,~$\alpha_4)=$(160,~0.5) \& (180,~0.9) which are consistent with \cite{Steiner,Buballa,Novikov,Baym,Bag,Benhar,Satz,Bhattacharya}. 
 
\begin{figure}[!ht]
\centering
\includegraphics[width=0.5\textwidth]{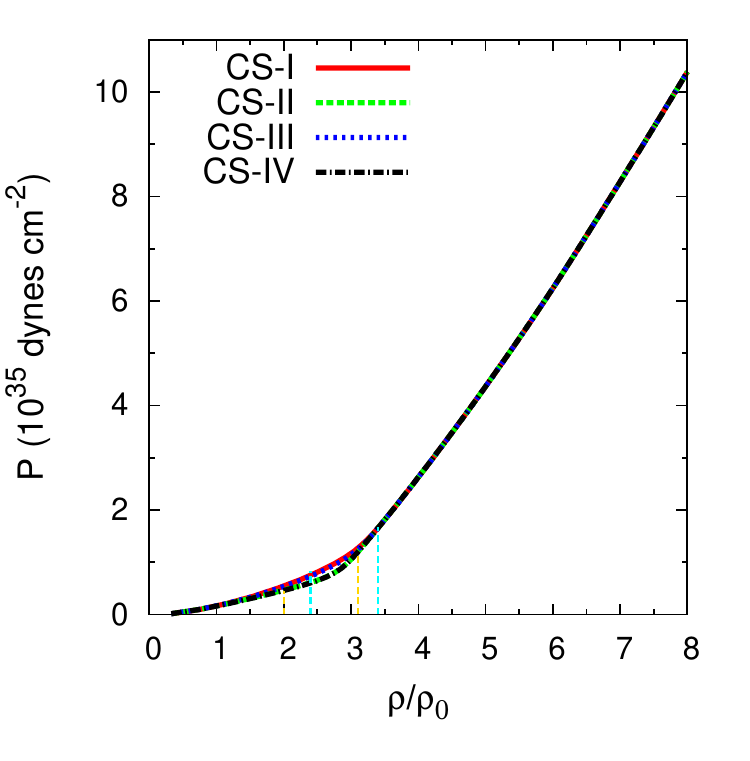}
\caption{\it Variation of pressure with baryon density for hybrid neutron star matter including $\Delta$ baryons and quarks for the different coupling schemes (as described in the text) and two different bag constant $B=(160~\rm{MeV})^4$ and interaction strength $\alpha_4=0.5$ with Gibbs Construction. The cyan vertical lines indicate the end of hadronic phase and beginning of quark phase for CS-I and CS-III while the yellow lines indicate same for CS-II and CS-IV.}
\protect\label{Prho_dq-1GC}
\end{figure}

\begin{figure}[!ht]
\centering
%\subfloat[With Gibbs Construction]
\includegraphics[width=0.5\textwidth]{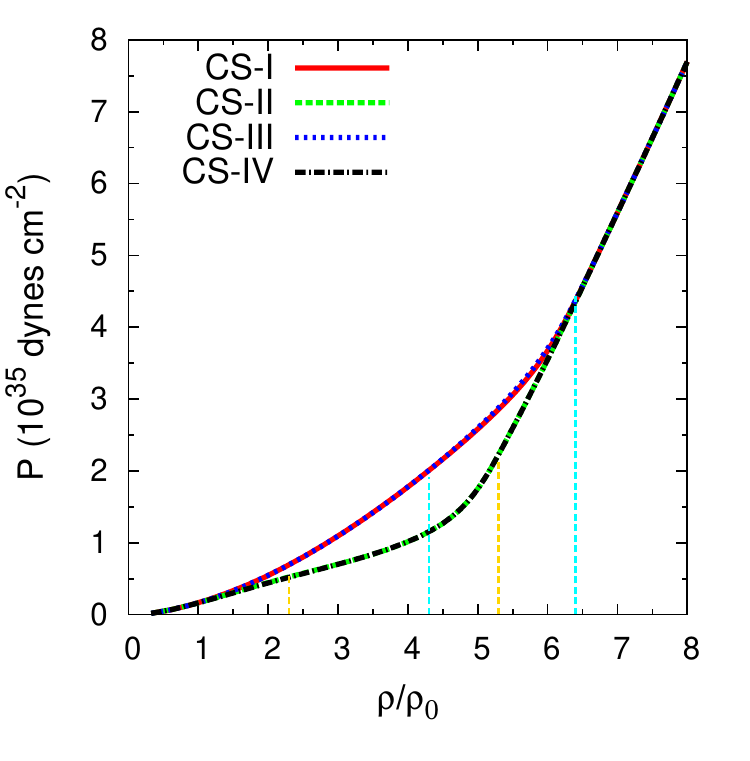}
\caption{\it Variation of pressure with baryon density for hybrid neutron star matter including $\Delta$ baryons and quarks for the different coupling schemes (as described in the text) and two different bag constant $B=(180~\rm{MeV})^4$ and interaction strength $\alpha_4=0.9$ with Gibbs Construction. The cyan vertical lines indicate the end of hadronic phase and beginning of quark phase for CS-I and CS-III while the yellow lines indicate same for CS-II and CS-IV.}
\protect\label{Prho_dq-2GC}
\end{figure}

\begin{figure}[!ht]
\centering
\includegraphics[width=0.5\textwidth]{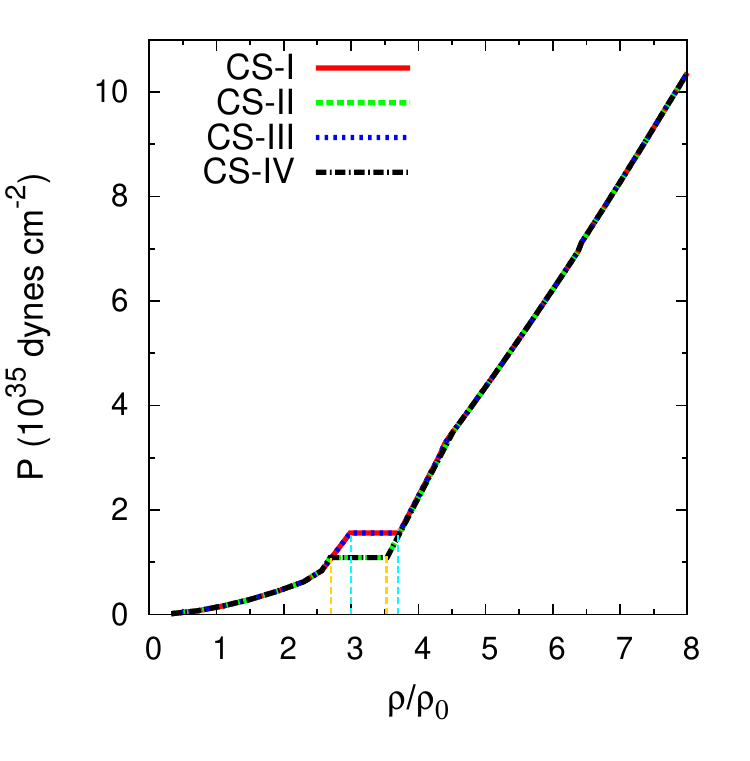}
\caption{Same as fig. \ref{Prho_dq-1GC} but with Maxwell Construction.}
\protect\label{Prho_dq-1MC}
\end{figure}

\begin{figure}[!ht]
\centering
%\subfloat[With Maxwell Construction]
\includegraphics[width=0.5\textwidth]{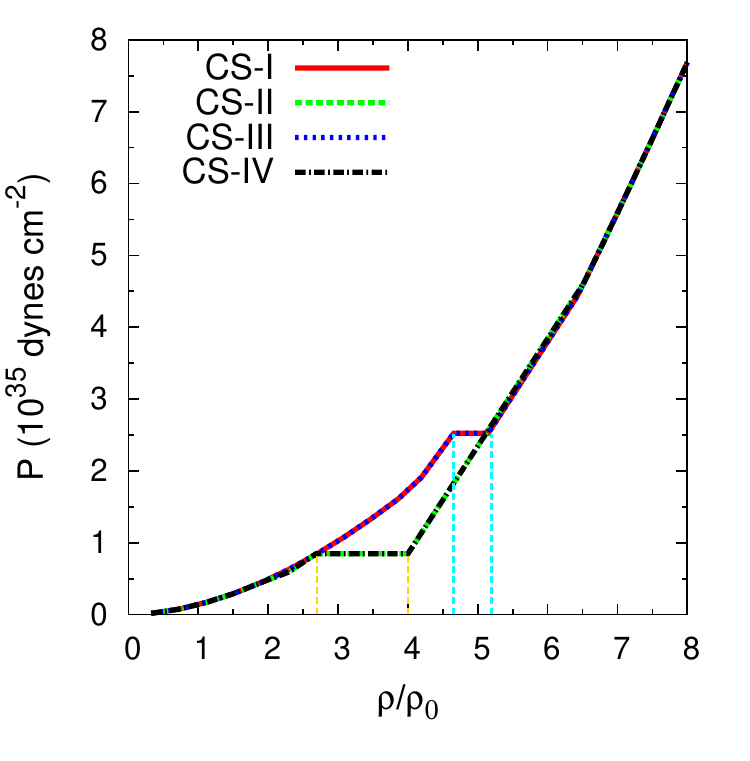}
\caption{Same as fig. \ref{Prho_dq-2GC} but with Maxwell Construction.}
\protect\label{Prho_dq-2MC}
%\protect\label{Prho_dq-2}
\end{figure}

 The hadron-quark crossover points in the $\mu-P$ plane shift to higher densities with the increase values of $B$ and $\alpha_4$. We plot the pressure as a function of baryon density with Gibbs construction (figs. \ref{Prho_dq-1GC} and \ref{Prho_dq-2GC}) and Maxwell construction (figs. \ref{Prho_dq-1MC} and \ref{Prho_dq-2MC}) for both the combinations of $B$ and $\alpha_4$. 
 
 As expected MC yields constant pressure over phase transition region unlike GC. Moreover, MC shows a much delayed phase transition compared to that obtained in case of GC. Our results are consistent with that of ref. \cite{Bhattacharya} where the hadronic phase is constructed with a rmf model while the pure quark phase has been constructed without considering the effects of strong repulsive quark interaction via $\alpha_4$.

 With GC, fig. \ref{Prho_dq-1GC} shows that for the first combination (smaller values of $B$ and $\alpha_4$) $B=\rm{(160~ MeV)^4}$ ; $\alpha_4=0.5$, the mixed phase initiates from 2.4$\rho_0$ and continues upto 3.4$\rho_0$ when $x_\rho = 0.5$ (CS-I and CS-III) while it persists from 2.0$\rho_0$ to 3.1$\rho_0$ for $x_\rho = 1.0$ (CS-II and CS-IV) for the same combination of $B$ and $\alpha_4$. For a higher value combination of both $B$ and $\alpha_4$ viz. $B=(180~ \rm{MeV)^4}$ ; $\alpha_4=0.9$, (fig. \ref{Prho_dq-2GC}) the mixed phase exists within density range (4.3 - 6.4)$\rho_0$ when $x_\rho = 0.5$ (CS-I and CS-III) and within (2.3 - 5.3)$\rho_0$ for $x_\rho = 1.0$ (CS-II and CS-IV). In case of MC, for the smaller combination of $B$ and $\alpha_4$ (fig. \ref{Prho_dq-1MC}), the hadronic phase extends upto density 3.0$\rho_0$ and pure quark phase begins at 3.7$\rho_0$ when $x_\rho = 0.5$ (CS-I and CS-III). For the same values of $B$ and $\alpha_4$ the hadronic phase is found upto 2.7 while the pure quark phase initiates at 3.5$\rho_0$ when $x_\rho = 1.0$ (CS-II and CS-IV). For higher combination of $B$ and $\alpha_4$, the hadronic phase is seen upto 4.6$\rho_0$ and pure quark phase begins at 5.1$\rho_0$ when $x_\rho = 0.5$ (CS-I and CS-III). When $x_\rho = 1.0$ (CS-II and CS-IV) the hadronic phase is seen upto 2.7$\rho_0$ while pure quark phase begins at 4.0$\rho_0$ (fig. \ref{Prho_dq-2MC}).
 
 We find that for a fixed value of bag constant and interaction strength, quark matter appears early for a lower value of iso-vector coupling ($x_\rho = 0.5$). The critical density of appearance of the $\Delta$s now shifts to higher densities and their concentration also decreases. However, there is still considerable percentage of $\Delta$s in NSM. For a fixed value of iso-vector coupling, both for GC and MC the increase in values of bag constant and interaction strength results in stiffer EoS resulting in more massive NS. 
 
\begin{figure}[!ht]
\centering
%\subfloat[With Gibbs Construction]
\includegraphics[width=0.48\textwidth]{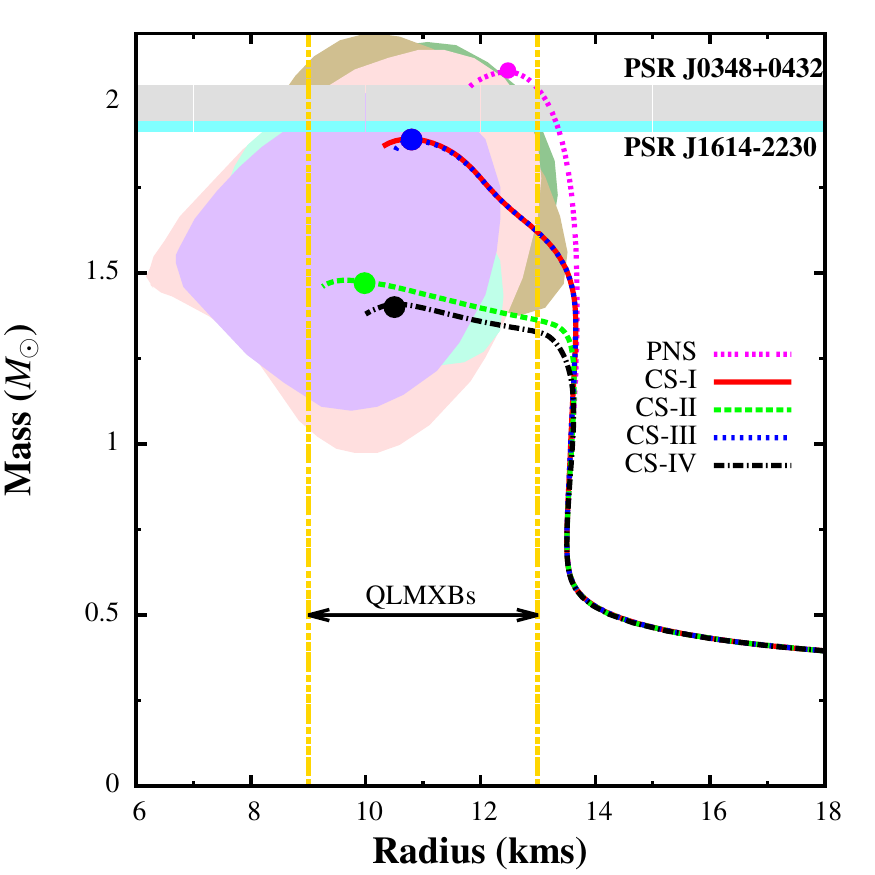}
\caption{\it Static neutron star gravitational mass-radius dependence for pure nucleon star (PNS) and hybrid neutron stars including $\Delta$ baryons and quarks for different coupling schemes and bag constant $B=(160~\rm{MeV})^4$ and interaction strength $\alpha_4=0.5$ for Gibbs Construction. Maximum mass limits imposed from recent observation of high mass stars PSR J1614-2230 ($M=(1.928 \pm 0.017) M_{\odot}$) \cite{Fonseca} and PSR J0348+0432 ($M=(2.01 \pm 0.04) M_{\odot}$) \cite{Ant} are also indicated. The region (marked with arrow) includes the estimation of radius within $(9 - 13)~km$ from spectral analysis of quiescent X-ray transients in low mass binaries (QLMXBs) \cite{fortin}. The areas enclosed by the dark green, golden, pink, aquamarine and purple curves indicate radius constraints from \cite{Ozel2} for sources 4U1820-30, 4U1724-207, KS1731-260, EXO1745-248 and 4U1608-52, respectively.}
\protect\label{mrdq-1GC}
\end{figure}

 The properties like central density, gravitational mass, baryonic mass and radius are obtained in static and spherical configurations of NS for lower and higher values of $B$ and $\alpha_4$ with both GC and MC. For lower $B$ and $\alpha_4$ ($B=(160~\rm{MeV)^4}$ ; $\alpha_4=0.5$) with GC the maximum values of gravitational mass $M$ (in $M_{\odot}$), with respective coupling schemes CS-I, CS-II, CS-III and CS-IV, are 1.89, 1.47, 1.89 and 1.40 for GC (fig. \ref{mrdq-1GC}).
 
\begin{figure}[!ht]
\centering
%\subfloat[With Gibbs Construction]
\includegraphics[width=0.48\textwidth]{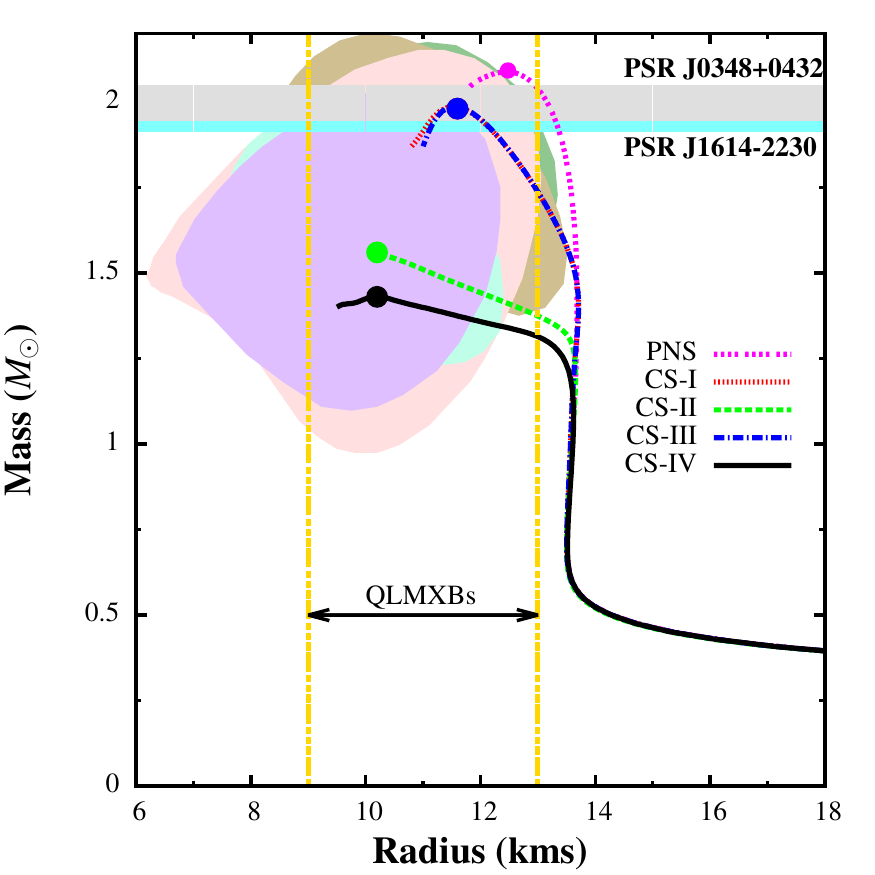}
\caption{\it Static neutron star gravitational mass-radius dependence for pure nucleon star (PNS) and hybrid neutron stars including $\Delta$ baryons and quarks for different coupling schemes and bag constant $B=(180~\rm{MeV})^4$ and interaction strength $\alpha_4=0.9$ for Gibbs Construction. Maximum mass limits imposed from recent observation of high mass stars PSR J1614-2230 ($M=(1.928 \pm 0.017) M_{\odot}$) \cite{Fonseca} and PSR J0348+0432 ($M=(2.01 \pm 0.04) M_{\odot}$) \cite{Ant} are also indicated. The region (marked with arrow) includes the estimation of radius within $(9 - 13)~km$ from spectral analysis of quiescent X-ray transients in low mass binaries (QLMXBs) \cite{fortin}. The areas enclosed by the dark green, golden, pink, aquamarine and purple curves indicate radius constraints from \cite{Ozel2} for sources 4U1820-30, 4U1724-207, KS1731-260, EXO1745-248 and 4U1608-52, respectively.}
\protect\label{mrdq-2GC}
\end{figure}

 The corresponding values of radius $R$ are 10.8 km, 9.9 km, 10.8 km and 10.5 km, respectively. For higher values of $B$ and $\alpha_4$ ($B=(180~\rm{MeV)^4}$ ; $\alpha_4=0.9$) for GC (fig. \ref{mrdq-2GC}) we have $M$ (in $M_{\odot}$), with respective coupling schemes CS-I, CS-II, CS-III and CS-IV as 1.98, 1.56, 1.98 and 1.47 with corresponding $R$ as 11.6 km, 10.2 km, 11.6 km and 10.2 km, respectively.
 
\begin{figure}[!ht]
\centering
\includegraphics[width=0.48\textwidth]{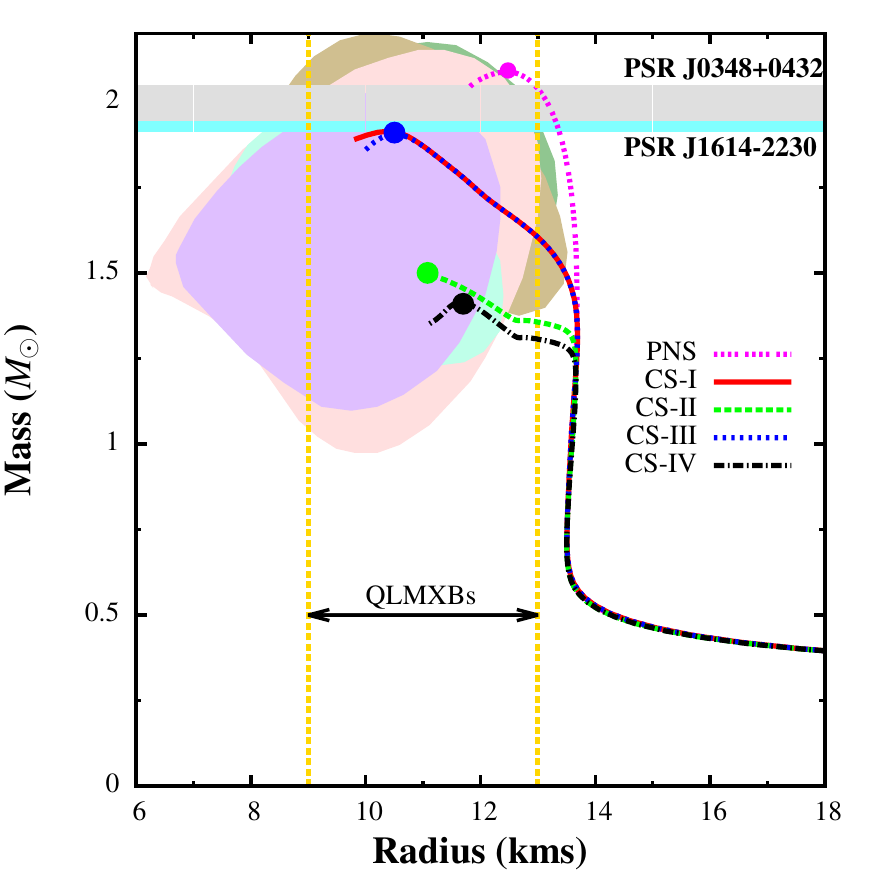}
\caption{Same as fig. \ref{mrdq-1GC} but for Maxwell Construction.}
\protect\label{mrdq-1MC}
\end{figure}

 In case of MC, lower values of $B$ and $\alpha_4$ yields maximum values of gravitational mass $M$ (in $M_{\odot}$), with respective coupling schemes CS-I, CS-II, CS-III and CS-IV as 1.91, 1.50, 1.91 and 1.47 (fig. \ref{mrdq-1MC}).

\begin{figure}[!ht]
\centering
\includegraphics[width=0.48\textwidth]{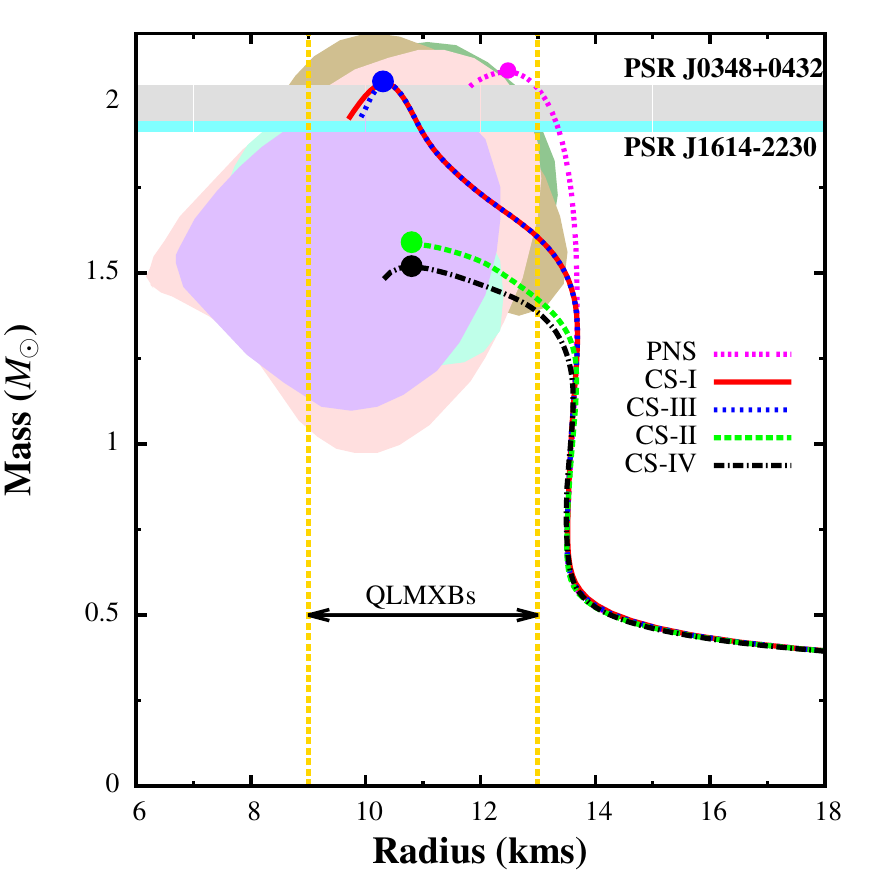}
\caption{\it Same as fig.\ref{mrdq-2GC} but for Maxwell Construction.}
\protect\label{mrdq-2MC}
\end{figure}

 The corresponding values of radius are 10.5 km, 11.1 km, 10.5 km and 11.7 km, respectively. With higher values of $B$ and $\alpha_4$, MC yields maximum gravitational mass $M$ (in $M_{\odot}$) as 2.06, 1.59, 2.06 and 1.52, respectively for coupling schemes CS-I, CS-II, CS-III and CS-IV at corresponding maximum radius 10.3 km, 10.8 km, 10.3 km and 10.8 km, respectively (fig. \ref{mrdq-2MC}).
 
 The overall results from hydrostatic equilibrium conditions are listed in table \ref{table-3} for the different coupling schemes and values of bag constant $B$ and repulsive strength $\alpha_4$ with Gibbs construction and Maxwell construction (in brackets).

\begin{table*}[ht!]
\centering
\tbl{Static neutron star properties for variation in the coupling schemes under consideration ie., $x_{i\Delta} = g_{i\Delta}/g_{iN}$, where $i=\sigma, \omega, \rho$ for two different bag constants ($B$) and interaction strengths ($\alpha_4$). The results from hydrostatic equilibrium conditions such as the central density of the star $\rho_c$ (in $\rho_0$), the gravitational mass $M$ (in $M_{\odot}$), the baryonic mass $M_B$ (in $M_{\odot}$), radius $R$ (km) and $R_{1.4}$ (km) are tabulated with Gibbs construction and Maxwell construction (in brackets).}
{\scriptsize{
\setlength{\tabcolsep}{3.5pt}
\begin{tabular}{ccccccccccccc}
\hline
\hline
\multicolumn{1}{c}{$B^{1/4}$}&
\multicolumn{1}{c}{$\alpha_4$}&
\multicolumn{1}{c}{Coupling Scheme}&
\multicolumn{1}{c}{$x_{\sigma\Delta}$}&
\multicolumn{1}{c}{$x_{\omega\Delta}$} &
\multicolumn{1}{c}{$x_{\rho\Delta}$} &
\multicolumn{1}{c}{$\rho_c$} &
\multicolumn{1}{c}{$M$} &
\multicolumn{1}{c}{$M_{B}$} & 
\multicolumn{1}{c}{$R$} & 
\multicolumn{1}{c}{$R_{1.4}$} &\\
\multicolumn{1}{c}{MeV} &
\multicolumn{1}{c}{} &
\multicolumn{1}{c}{} &
\multicolumn{1}{c}{} &
\multicolumn{1}{c}{} &
\multicolumn{1}{c}{} &
\multicolumn{1}{c}{$(\rho_0)$} &
\multicolumn{1}{c}{($M_{\odot}$)} &
\multicolumn{1}{c}{($M_{\odot}$)} &
\multicolumn{1}{c}{($km$)} & 
\multicolumn{1}{c}{($km$)} \\
\hline
160  &0.5  &CS-I  &1.35  &1.0   &1.0   &5.3(7.0)  &1.89(1.91)  &1.98(1.99)   &10.8(10.5) &13.4(13.4)\\
     &    &CS-II  &     &      &0.5   &6.9(7.2)  &1.47(1.50)  &1.65(1.68)   &9.9(11.1) &12.0(12.3)\\ 
     &    &CS-III &1.20 &0.8   &1.0   &5.3(7.0)  &1.89(1.91)  &1.98(1.99)   &10.8(10.5) &13.4(13.4)\\
     &    &CS-IV  &     &      &0.5   &5.9(6.4)  &1.40(1.47)  &1.57(1.60)   &10.5(11.7) &10.5(11.9)\\ 
\hline
180  &0.9  &CS-I   &1.35   &1.0   &1.0   &5.4(7.0)  &1.98(2.06)  &2.10(2.15) &11.6(10.3) &13.4(13.4)\\
     &    &CS-II  &     &      &0.5   &6.0(6.8)  &1.56(1.59)  &1.72(1.78) &10.2(10.8) &12.5(13.1)\\ 
     &    &CS-III &1.20  &0.8   &1.0  &5.4(7.0)  &1.98(2.06)  &2.10(2.15) &11.6(10.3) &13.4(13.4)\\
     &    &CS-IV  &     &      &0.5   &6.9(6.7)  &1.47(1.52)  &1.65(1.68) &10.2(10.8) &11.1(13.0)\\ 
\hline
\hline
\end{tabular}
}}
%\end{center}
\protect\label{table-3}
\end{table*}

 As expected, the inclusion of quark matter stiffens the EoS and increases the maximum gravitational mass. Consistent with results of works like \cite{Bhattacharya} with different approaches, MC yields more massive configurations of NS due to delayed appearance of quarks compared to that in case of GC. We find that for a fixed value of $B$ and $\alpha_4$, the variation of the scalar and vector couplings do not bring any significant change to the gross properties of hybrid stars like central density, the maximum gravitational mass, the baryonic mass and the radius. It is still the iso-vector coupling $x_{\rho}$ alone which bring substantial change in these quantities if $B$ and $\alpha_4$ are kept constant. With fixed values of $\Delta$ couplings, the variation of bag constant and repulsive interactions of the quarks have profound influence on the properties of hybrid neutron stars. With both GC and MC, the increase in $B$ and $\alpha_4$ leads to massive NS configurations and fulfill the high gravitational mass ($M=(2.01 \pm 0.04) M_{\odot}$) constraint \cite{Ant} with same $\Delta$ couplings used before. Moreover, with the inclusion of quark matter, the radii obtained for the hybrid star for all the couplings and values of $B$ and $\alpha_4$ are still within the limits imposed from observational analysis of QLMBXs \cite{fortin} and canonical NS configurations \cite{latt}. The estimates also satisfy constraint on radius from \cite{Ozel2}. Also all the values of $R_{1.4}$ obtained with the hybrid star configurations for all the couplings and values of $B$ and $\alpha_4$ agree with the recent bounds specified from the observation of gravitational wave (GW170817) from BNS merger \cite{Abbot,Fattoyev,Most,De}. There is also noticeable increase in the central density on including quark matter (table \ref{table-3}). 
 
 Interestingly, with our hybrid EoS we could also satisfy the constraint on baryonic mass from PSR J0737-3039 B \cite{Podsiadlowski}. The pulsar PSR J0737-3039 B, with maximum gravitational mass $M_G=(1.249 \pm 0.001) M_{\odot}$ \cite{Burgay} has maximum baryonic mass $M_B=(1.366 - 1.375)M_{\odot}$ \cite{Podsiadlowski} assuming that the pulsar was formed from an electron capture supernova. 

\begin{figure}[ht!]
\centering
%\subfloat[With Gibbs Construction]
\includegraphics[width=0.5\textwidth]{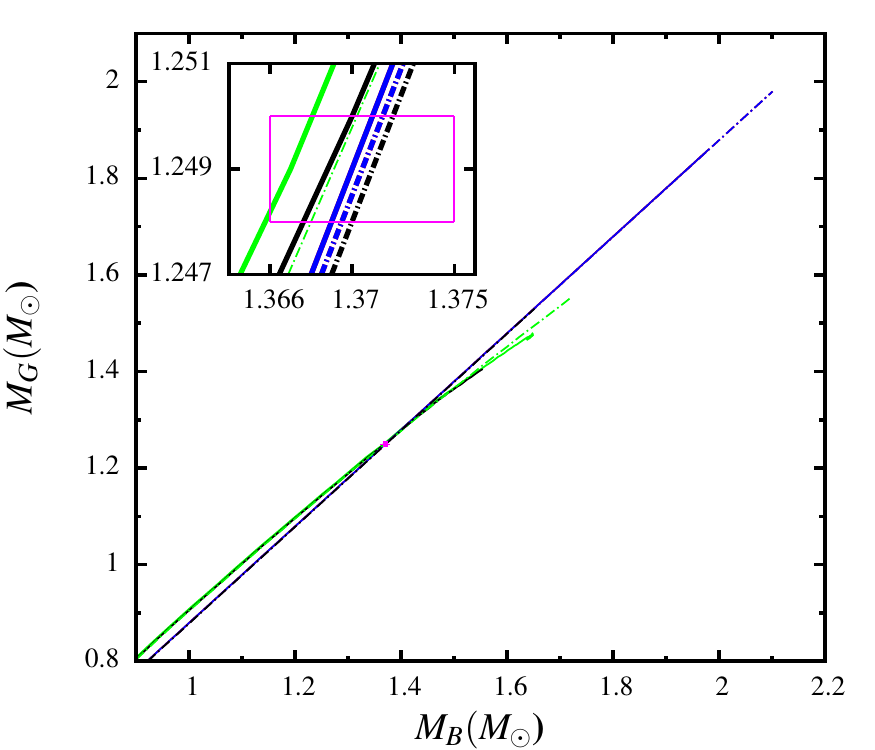}
\caption{\it Baryonic mass ($M_B$) versus gravitational mass ($M_G$) for static hybrid neutron star with Gibbs Construction. The solid lines represent $B=(160~\rm{MeV})^4$ ; $\alpha_4=0.5$ configuration while dotted lines represent $B=(180~\rm{MeV})^4$ ; $\alpha_4=0.9$ configuration. The different colors represent same coupling schemes as figs. \ref{mrdq-1GC},\ref{mrdq-1MC} and \ref{mrdq-2GC}, \ref{mrdq-2MC}. The magenta box represent the constraint of ref.\cite{Podsiadlowski} on baryonic mass ($M_B=(1.366 - 1.375)M_{\odot}$) for Pulsar B of binary system PSR J0737-3039 with gravitational mass ($M_G=(1.249 \pm 0.001) M_{\odot}$) \cite{Burgay}.}
\protect\label{mg_mbGC}
\end{figure}

%\hfill
\begin{figure}[ht!]
\centering
\includegraphics[width=0.5\textwidth]{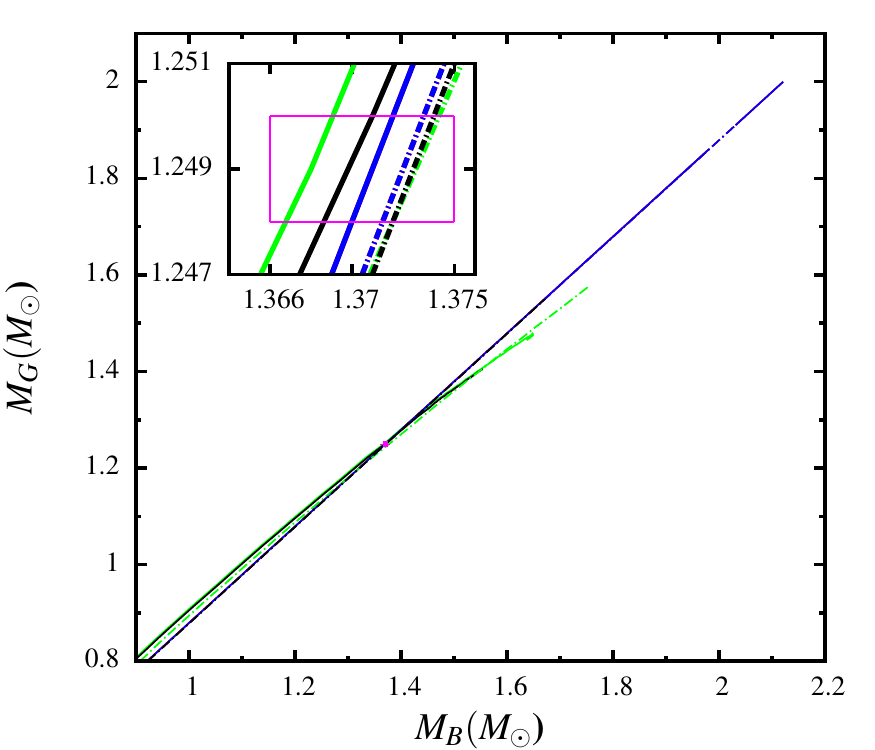}
\caption{Same as fig. \ref{mg_mbGC} but for Maxwell Construction.}
\protect\label{mg_mbMC}

%\protect\label{mg_mb}
\end{figure}

 We plot the baryonic mass $M_B$ versus the gravitational mass $M_G$ for both $B=(160~\rm{MeV)^4}$ ; $\alpha_4=0.5$ and $B=(180~\rm{MeV)^4}$ ; $\alpha_4=0.9$ for GC (fig. \ref{mg_mbGC}) and MC (fig. \ref{mg_mbMC}). The inserts of these figures show that the constraint on baryonic mass \cite{Podsiadlowski} is successfully satisfied with all the coupling schemes for both $B=(160~\rm{MeV)^4}$ ; $\alpha_4=0.5$ and $B=(180~\rm{MeV)^4}$ ; $\alpha_4=0.9$ both in case of GC and MC. For all the cases the value of $M_B$ for $M_G=(1.249 \pm 0.001) M_{\odot}$ lie within the range $M_B=(1.366 - 1.375)M_{\odot}$. In this work we find that this constraint from PSR J0737-3039 \cite{Podsiadlowski} is fulfilled (figs. \ref{mg_mbGC}, \ref{mg_mbMC}) with our hybrid EoS. 
 
  Overall, within the framework of general hydrostatic equilibrium based on general theory of relativity (GTR), the transition of hadronic matter to quark matter leads to more massive and compact hybrid star configurations compared to that obtained with only $\Delta$s and nucleons. In this work we aim to investigate the role of $\Delta$ baryons in determining the properties of NS and it is well-known that the formation of hyperons suppresses the formation of $\Delta$s considerably in NSM. Therefore like many other works \cite{prc92,Zhou,Oliveira3,Hu,Chen} we do not consider the contribution of hyperons in the pure hadronic part of EoS in order to maintain similarity of quark structure in hadronic phase. However, one may include hyperons in the pure hadronic matter. In general the inclusion of hyperons reduces the maximum mass of NS than in case of pure neucleonic star \cite{Miyatsu,Stone,Dhiman,Dexheimer,Bednarek,Bednarek2, Weissenborn12,Weissenborn14,Agrawal,Lopes, Oertel,Colucci,Dalen,Lim,Rabhi,TKJ2,TKJ3,Sen}. In such cases, the choice of bag constant $B$ and interaction strength $\alpha_4$ should be made accordingly in order to meet with the $2 M_{\odot}$ mass criteria of NS.

%\newpage
\section{Conclusion}
We present a comprehensive aspect of formation of delta baryons in NSM and the resulting NS properties in an effective chiral model in mean-field approach. The model has been tested earlier to study nuclear matter properties and the parameter set considered is well consistent with saturated nuclear matter properties and heavy-ion collision data. This work is particularly aimed to investigate the effect of varied delta-meson couplings on the EoS and consequent transition of hadronic matter to quark matter in order to satisfy the maximum mass constraint on NS, imposed from recent observations of pulsars PSR J1614-2230 and PSR J0348+0432. In absence of concrete experimental data, the delta-meson couplings are chosen consistently with that prescribed from finite density QCD calculations \cite{qcd,qcd1}. We find that considerable amount of $\Delta$ particles, particularly $\Delta^-$ and $\Delta^0$ can be formed in dense NSM and the corresponding NS properties are very sensitive to the iso-vector couplings (\ref{app_sensitivity}). Overall, early appearance of $\Delta$s in matter results in very compact stellar configurations. The radius predicted with inclusion of $\Delta$s, are in excellent agreement with recent limits imposed on the NS radius estimates from QLMXBs. The obtained values of $R_{1.4}$ are in agreement with the recent bounds specified from the observation of gravitational wave (GW170817) from BNS merger. However, with only nucleons and deltas, we could meet the maximum mass constraint for none of the couplings and therefore we looked for the possible phase transition from hadronic matter to quark matter using both Gibbs and Maxwell constructions. With the hybrid EoS, we could satisfy the aforesaid maximum mass constraint with bag constant $B=(180~\rm{MeV)^4}$ and interaction strength $\alpha_4=0.9$ for models CS-I and CS-III as described in the text with both GC and MC approaches within the framework of general hydrostatic equilibrium based on GTR. By including quarks, the maximum mass of the star increases by $\approx (6-10)\%$ from that obtained with pure hadronic matter without affecting the radius much. Thus the maximum mass constraint could be satisfied only when we allow the non-strange hadronic matter including $\Delta$s to undergo phase transition to quark matter at relevant high densities, forming hybrid stars. Such phase transitions not only satisfy the maximum mass criterion but also make the radius and the baryonic mass estimates within the limits imposed from observational analysis of low-mass binaries. We also look for and discuss briefly about the possibility of forming strange quark star from neutron (hadronic) star in \ref{app_mgmbB}. We need to improve on the symmetry energy aspects of the present model, which may play crucial role in high density asymmetric matter and NS properties.

\appendix

\section{Sensitivity of formation of $\Delta$s in NSM \& Neutron Star properties to scalar, vector \& iso-vector couplings}
\label{app_sensitivity}

\begin{figure}[!ht]
\centering
{\includegraphics[width=1.0\textwidth]{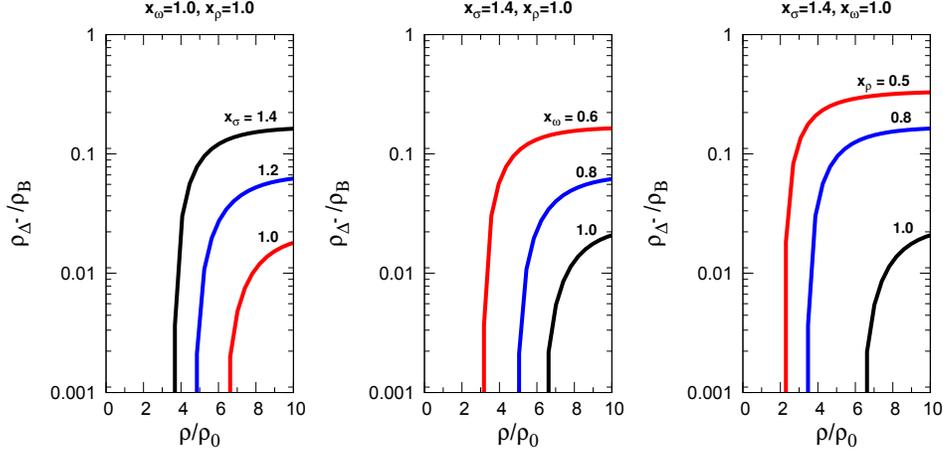}
\caption{Population of $\Delta^-$ baryons as a function of normalized baryon density for different values of $x_{\sigma}$, $x_{\omega}$ and $x_{\rho}$.}
\protect\label{sensitivity}}
\end{figure}

In order to test the sensitivity of the couplings on the appearance and concentration of $\Delta$s and the resultant NS properties, we vary the couplings one at a time. The values of $x_{\sigma}$ and $x_{\omega}$ are within the prescribed range of \cite{qcd} and $x_{\rho}$ is varied accordingly. In fig. \ref{sensitivity}, we show the normalized population of $\Delta^-$ as a function of baryon density with varying $x_{\sigma}$, $x_{\omega}$ and $x_{\rho}$ individually. In the left panel we have varied the scalar coupling keeping the vector and iso-vector couplings constant. Similarly in the middle and right panels the vector and iso-vector coupling is varied keeping the other two fixed.

 We find from fig. \ref{sensitivity} that the early appearance of $\Delta^-$ with increased concentration is favored with the increase of nucleon-delta attractive strength via $x_{\sigma}$ (left panel). However, their appearance is relatively late and their population decreases since the increase in $x_{\omega}$ (middle panel) increases the repulsive strength between the deltas and nucleons \cite{Oliveira,Oliveira2,DragoPRC}. However, the effect is most pronounced in case where we vary $x_{\rho}$ (right panel) \cite{prc92}. For example, $\Delta^-$ appears as early as at $2.1\rho_0$ with highest concentration when $x_{\rho}=0.5$ and at $6.8\rho_0$ when $x_{\rho}=1.0$. With the decrease of $x_{\rho}$, the concentration of $\Delta^-$ also increases. This is because the iso-vector coupling strength is largely affects the symmetry energy aspects of neutron stars which in turn are strongly co-related to the critical density and concentration of $\Delta$s \cite{prc92}. With the variations in couplings mentioned above, we then calculate the resulting NS properties like mass and radius along with the corresponding delta potential (using eq. \ref{potdPNM}). The results are tabulated in table \ref{sensitivity-table}. Independent variations of $x_{\sigma}$ and $x_{\omega}$ also show changes in properties of NS but they are most sensitive to $x_{\rho}$. However, it should be kept in mind that unlike that of $x_{\rho}$, the variation of $x_{\sigma}$ and $x_{\omega}$ are constrained within the range prescribed by \cite{qcd}. At present constraints on variation of $x_{\rho}$ are still uncertain in the existing literatures from both theoretical and experimental perspectives. 

\begin{table*}[ht!]
\centering
\tbl{Critical densities for appearance of $\Delta^-$ ($\rho_{\Delta^-}^{crit}$), $\Delta^-$ potential $U_{\Delta^-}$ (in MeV) and static neutron star properties such as gravitational mass $M$ (in $M_{\odot}$), radius $R$ (km) and $R_{1.4}$ (km) are tabulated for variation in the couplings ie., $x_{i\Delta} = g_{i\Delta}/g_{iN}$, where $i=\sigma, \omega, \rho$.}
{\small{
\setlength{\tabcolsep}{5.5pt}
\begin{tabular}{ccccccccccccc}
\hline
\hline
\multicolumn{1}{c}{$x_{\sigma}$}&
\multicolumn{1}{c}{$x_{\omega}$} &
\multicolumn{1}{c}{$x_{\rho}$} &
\multicolumn{1}{c}{$U_{\Delta^-}$} &
\multicolumn{1}{c}{$\rho_{\Delta^-}^{crit}$} &
\multicolumn{1}{c}{$M$} &
\multicolumn{1}{c}{$R$} & 
\multicolumn{1}{c}{$R_{1.4}$} &\\
\multicolumn{1}{c}{} &
\multicolumn{1}{c}{} &
\multicolumn{1}{c}{} &
\multicolumn{1}{c}{(MeV)} &
\multicolumn{1}{c}{($\rho_0$)} &
\multicolumn{1}{c}{($M_{\odot}$)} &
\multicolumn{1}{c}{($km$)} & 
\multicolumn{1}{c}{($km$)} &\\
\hline

1.0  &1.0 &1.0 &39.0  &6.8 &1.75 &12.4 &13.4\\
1.2  &    &    &10.8  &6.8 &1.70 &12.4 &13.4\\
1.4  &    &    &-17.4 &6.8 &1.67 &12.4 &13.4\\
\hline
1.4  &0.6 &1.0 &-34.4 &5.2 &1.66 &12.5 &13.4\\
     &0.8 &    &-22.4 &5.8 &1.71 &12.4 &13.4\\
     &1.0 &    &-17.4 &6.8 &1.75 &12.4 &13.4\\
\hline     
1.4  &1.0 &0.5 &-70.3 &2.1 &1.40 &11.7 &12.7\\
     &    &0.8 &-34.3 &3.5 &1.69 &12.6 &13.4\\     
     &    &1.0 &-17.4 &6.8 &1.75 &12.4 &13.4\\
\hline
\hline
\end{tabular}
}}
%\end{center}
\protect\label{sensitivity-table}
\end{table*}

\vspace*{-0.5cm}
\section{Possibility of formation of Strange Quark Stars from Neutron Stars}
\label{app_mgmbB}

\begin{figure}[ht!]
\centering
%\subfloat[With Gibbs Construction]
\includegraphics[width=0.48\textwidth]{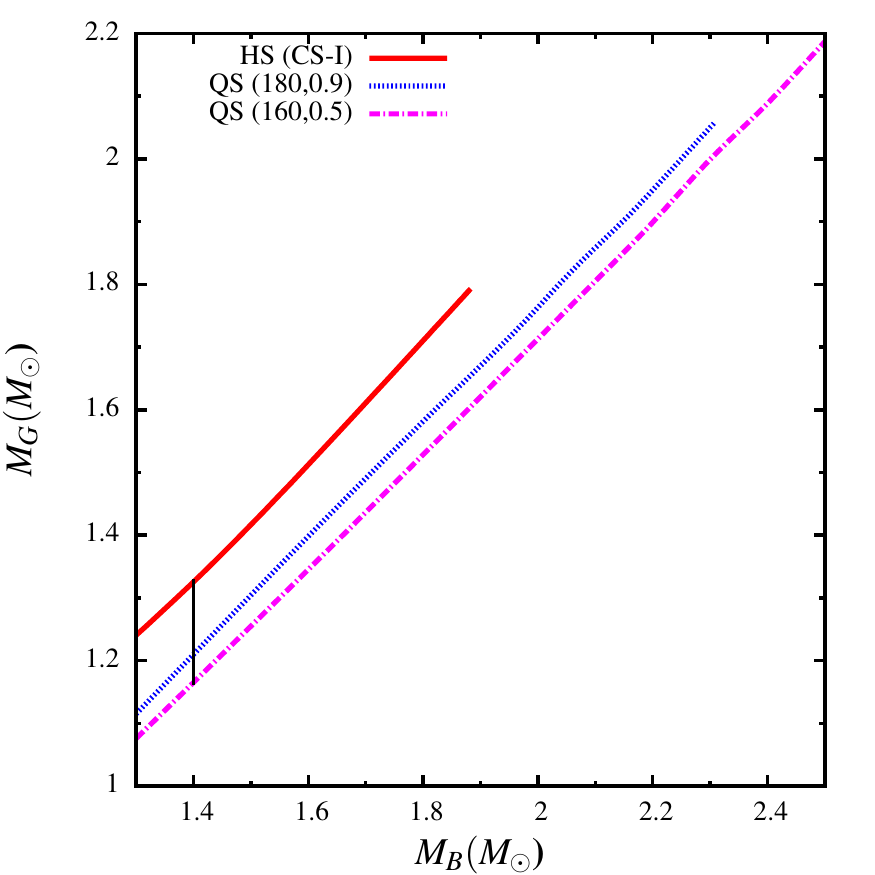}
\caption{\it Baryonic mass ($M_B$) versus gravitational mass ($M_G$) for static hadronic star including the $\Delta$s (HS) and strange quark stars (QS) with the MIT Bag model having parameters ($B^{1/4},\alpha_4$)=(160,0.5),(180,0.9). The black vertical line indicates the values of $M_G$ for HS and QSs for $M_B=1.4 M_{\odot}$.}
\protect\label{mgmbB}
\end{figure}

We now investigate the possibility of conversion of neutron stars into strange quark stars. As suggested in literature, there are several possible ways in which hadronic matter can undergo total conversion to strange quark matter viz. clustering of $\Lambda$ hyperons, kaon condensation, burning of hadrons to strange quark matter and seeding from outside through accretion \cite{Olesen,HChen,Alcock,Dragoq1,Dragoq2,Bom2000,Bombaci16,Log12,Drago15}. Such conversions costs emission of huge amount of energy, with subsequent gamma ray or neutrino bursts \cite{Cheng,Bom2000,Berezhiani,Rujula}. In this work we investigate the possibility of formation of strange quark stars from neutron stars in terms of energy with conservation of baryonic mass \cite{Dragoq1,Dragoq2,HChen,Bom2000}. We plot in fig. \ref{mgmbB} the baryonic mass versus the gravitational mass for hadronic star including the $\Delta$s (HS) and strange quark star (QS) with the MIT Bag model having parameters ($B^{1/4},\alpha_4$)=(160,0.5),(180,0.9).

 For HS we choose the configuration with delta coupling scheme CS-I or CS-III since both yield the same stellar configuration (fig. \ref{mrd} and table \ref{table-2}) due to almost overlapping EoS (fig. \ref{eos_d}). It is seen from \ref{mgmbB} that for a fixed hadronic star configuration, the quark star with the same baryon mass has a smaller gravitational mass. For example we show in fig. \ref{mgmbB} (with the black vertical line) that for fixed $M_B=1.4 M_{\odot}$, the corresponding gravitational mass for hadronic configuration is $1.33 M_{\odot}$ and for strange quark star configuration the gravitational mass is $1.21 M_{\odot}$ for $B=\rm{(180~ MeV)^4}$, $\alpha_4=0.9$ and $1.17 M_{\odot}$ for $B=\rm{(160~ MeV)^4}$, $\alpha_4=0.5$. Therefore the formation of strange quark stars from hadron star is supported in terms of energy \cite{Dragoq1,Dragoq2,HChen,Bom2000}. The gravitational binding energy $E_G$ for any configuration is given by

\begin{eqnarray}
E_G = (M_B - M_G)~c^2
\end{eqnarray}

The energy $E_r$ released in the process of conversion of hadronic star into strange quark star can be interpreted as difference in gravitational binding energy of the strange quark star $E^{QS}_G$ and the hadronic star $E^{HS}_G$ with the baryonic mass conserved \cite{HChen,Bom2000}.

\begin{eqnarray}
E_r \equiv E^{QS}_G - E^{HS}_G
\end{eqnarray}

We find that for a hadronic star with gravitational mass $M_G=1.4 M_{\odot}$, the energy released $E_r$ is $2.19 \times 10^{53}$ erg for $B=\rm{(180~ MeV)^4}$, $\alpha_4=0.9$ and $2.93 \times 10^{53}$ erg for $B=\rm{(160~ MeV)^4}$, $\alpha_4=0.5$. Our estimates of $E_r$ for a $1.4 M_{\odot}$ star are quite consistent with the range specified for the same by other works like \cite{HChen,Bom2000}. Consistent with results of works like \cite{HChen} we also find that more energy is released in the conversion process as the gravitational mass of hadronic star increases. The maximum energy released for maximum gravitational mass of hadronic star $M_G=1.8 M_{\odot}$ is $2.51 \times 10^{53}$ erg for $B=\rm{(180~ MeV)^4}$, $\alpha_4=0.9$ and $3.40 \times 10^{53}$ erg for $B=\rm{(160~ MeV)^4}$, $\alpha_4=0.5$. However, it is not possible to comment on whether such conversions can take place in case of massive neutron stars like PSR J1614-2230 and PSR J0348+0432 because the maximum mass of the pure hadronic star, obtained with our model is quite less than the mass of these pulsars.

\end{document}